\documentclass[aps, prd, onecolumn,showpacs, showkeys, preprintnumbers, preprint, nofootinbib, superscriptaddress]{revtex4-2}
\usepackage[sort&compress]{natbib}
\usepackage{multirow}
\usepackage{graphicx}
\usepackage{amssymb,amsbsy,amsmath,amsfonts}
\usepackage[latin1]{inputenc}

\begin{document}
\title{Partial decay widths of $\phi(2170)$ to kaonic resonances}
\author{Brenda B. Malabarba}
\email{brenda@if.usp.br}
\affiliation{Universidade de Sao Paulo, Instituto de Fisica, C.P. 05389-970, Sao 
Paulo, Brazil.}

\author{Xiu-Lei Ren}
\email{xiulei.ren@uni-mainz.de}
\affiliation{Institut f\"ur Kernphysik $\&$ Cluster of Excelence PRISMA${}^+$, Johannes Gutenberg-Universit\"at Mainz, D-55099 Mainz, Germany.}

\author{K. P. Khemchandani}
\email{kanchan.khemchandani@unifesp.br}
\affiliation{Universidade Federal de Sao Paulo, C.P. 01302-907, Sao Paulo, Brazil.}

\author{A. Mart\'inez Torres}
\email{amartine@if.usp.br}
\affiliation{Universidade de Sao Paulo, Instituto de Fisica, C.P. 05389-970, Sao 
Paulo, Brazil.}

\begin{abstract}
In this work we study the strong decays of $\phi(2170)$ to final states involving the kaonic resonances $K(1460)$, $K_1(1270)$ and $K_1(1400)$, on  which experimental data have recently been extracted by the  BESIII Collaboration.  The formalism developed here is based on interpreting $\phi(2170)$ and $K(1460)$ as states arising from three-hadron dynamics, which is inspired by our earlier works. For   $K_1(1270)$ and $K_1(1400)$ we investigate different descriptions, such as a mixture of states belonging to the nonet of axial resonances, or the former one as a state originating from the vector-pseudoscalar dynamics.  The ratios among the partial widths of $K^+(1460)K^-$, $K^+_1(1400)K^-$ and $K^+_1(1270)K^-$ obtained are compatible with the experimental results, reinforcing the three-body nature of $\phi(2170)$.  Within our formalism, we can also explain the suppressed decay of $\phi(2170)$ to $K^*(892) \bar K^*(892)$,  as found by the BESIII Collaboration. Furthermore, our results can be useful in clarifying the properties of $K(1460)$, $K_1(1270)$ and $K_1(1400)$ when higher statistics data would be available.
\end{abstract}



\maketitle
\date{\today}

\section{Introduction}
The BESIII Collaboration has recently~\cite{Ablikim:2020pgw} studied some properties of $\phi(2170)$~\cite{Aubert:2006bu,Aubert:2007ur,Ablikim:2007ab,Shen:2009zze,Lees:2011zi} via the process $e^+e^-\to K^+ K^- \pi^0 \pi^0$, where a signal with a mass of $2126.5\pm16.8\pm12.4$ MeV and a width of $106.9\pm32.1\pm28.1$ MeV is observed and identified with $\phi(2170)$. The cross sections for different configurations of the final state $K^+ K^- \pi^0 \pi^0$ are obtained and the product $\mathcal{B}r\Gamma^{e^+e^-}_R$ is determined, where $\Gamma^{e^+e^-}_R$ corresponds to the partial decay width of $\phi(2170)$ to $e^+ e^-$ and $\mathcal{B}r$ is the branching fraction of $\phi(2170)$ to a specific configuration of the final state. In particular, the decay channels $K^+(1460) K^-$, $K^+_1(1400) K^-$, $K^{*+}(1410) K^-$, $K^+_1(1270)K^-$, and $K^{*+}(892) K^{*-}(892)$ are investigated. To determine $\mathcal{B}r\Gamma^{e^+e^-}_R$ in Ref.~\cite{Ablikim:2020pgw}, the data are fitted under the assumption that the signal observed for $\phi(2170)$ in the different decay channels should have the same mass and width. While a peak or a dip which can be related to $\phi(2170)$ is seen in the cross sections of $e^+ e^-\to K^+(1460)K^-,~ K^+_1(1400) K^-,~K^+_1(1270) K^-$, no evident peak/dip for $\phi(2170)$ is observed in the cross section for $e^+ e^-\to K^{*+}(892) K^{*-}(892)$~\cite{Ablikim:2020pgw}. Also, the decay to $K^{*+}(1410) K^-$ is found to have a statistical significance less than 3$\sigma$. In view of these results, it is concluded in Ref.~\cite{Ablikim:2020pgw} that if the signal observed in the process $e^+e^-\to K^+ K^- \pi^0 \pi^0$ is a manifestation of $\phi(2170)$, the decay of this state via $K^{*+}(892) K^{*-}(892)$ and via $K^{*+}(1410) K^-$ is suppressed as compared to the other three modes, i.e., $K^+(1460) K^-$, $K^+_1(1400) K^-$, $K^+_1(1270)K^-$.

In this work, we are going to determine the partial decay widths of $\phi(2170)$ to the channels $K^+(1460) K^-$, $K^+_1(1400) K^-$, $K^+_1(1270)K^-$ and compare their ratios with the experimental results obtained in Ref.~\cite{Ablikim:2020pgw}. These decay widths depend on the nature of the states involved, i.e., $\phi(2170)$, $K^+(1460)$, $K^+_1(1400)$ and $K^+_1(1270)$, and several theoretical models considering them as standard quark-antiquark states, tetraquarks, hadrons molecules or hybrid states have been proposed in the recent years [see, for example, Refs.~\cite{Barnes:1996ff,Ding:2006ya,Wang:2006ri,Ding:2007pc,MartinezTorres:2008gy,Drenska:2008gr,AlvarezRuso:2009xn,MartinezTorres:2010ax,Ho:2019org,Agaev:2019coa,Agaev:2020zad} for $\phi(2170)$, Refs.~\cite{Godfrey:1985xj,Albaladejo:2010tj,Torres:2011jt,Kezerashvili:2015aca,Shinmura:2019nqw,Filikhin:2020ksv} for $K(1460)$, and Refs.~\cite{Palomar:2003rb,Lutz:2003fm,Roca:2005nm,Geng:2006yb,Zhou:2014ila} for $K_1(1270)$ and $K_1(1400)$]. It has been discussed in Ref.~\cite{Ablikim:2020pgw} that the experimental findings on the decays of $\phi(2170)$ are incompatible with the predictions of the models considering a $s\bar s$ or hybrid description for it. Indeed, a $3{}^3 S_1$ $s\bar s$ description~\cite{Barnes:2002mu} (where the spectroscopy notation $n{}^{2S+1}L_J$ is used to denote the n$th$ state with total angular momentum $J$, spin $S$ and orbital angular momentum $L$) leads to a large width for $\phi(2170)$, $\sim 300$ MeV, which is not compatible with the experimental findings. Within a different quantum number attribution to the $s\bar s$ system, treating $\phi(2170)$ as a $2 {}^3 D_1$ state, the partial decay widths of $\phi(2170)$ to different channels were investigated  in Ref.~\cite{Ding:2007pc}. Within such a model $\phi(2170)$ has a larger decay width to $K^*(892)\bar K^*(892)$ and $K^*(1410)\bar K$ than to channels like $K(1460) \bar K$, $K_1(1400)\bar K$ and $K_1(1270)\bar K$. Such a decay pattern does not seem to be compatible with the findings of Ref.~\cite{Ablikim:2020pgw}. A different nature for $\phi(2170)$, a hybrid $s\bar s g$ state, was proposed in Ref.~\cite{Ding:2006ya}. According to the calculations performed in Refs.~\cite{Ding:2006ya,Ding:2007pc}, the  partial decay width of $\phi(2170)$ to $K^*(1410)\bar K$ is larger, or of similar order, as compared to the corresponding value for $K_1(1270)\bar K$, with the mode $K(1460) \bar K$ forbidden for decay due to a spin selection rule~\cite{Page:1998gz}. These properties of $\phi(2170)$ appear to be in disagreement with those found in Ref.~\cite{Ablikim:2020pgw}, as mentioned by the BESIII Collaboration.  
Also, such a hybrid interpretation for the internal structure of $\phi(2170)$ seems not to be supported by Lattice QCD studies~\cite{Dudek:2011bn} and QCD Gaussian sum rules calculations~\cite{Ho:2019org}. In case of a tetraquark nature assigned to $\phi(2170)$~\cite{Wang:2006ri,Drenska:2008gr,Deng:2010zzd}, a difficulty in obtaining a mass compatible with the experimental data has been reported in Ref.~\cite{Wang:2006ri} while using standard QCD sum rules. Though no predictions are available for the decay widths to the channels studied by BESIII~\cite{Ablikim:2020pgw} within a tetraquark model for $\phi(2170)$, it has been argued in Ref.~\cite{Ding:2007pc} that such an interpretation would imply a dominant decay to $\phi\eta(\eta^\prime)$, which cannot be inferred from the experimental findings~\cite{PDG}.

Therefore, the properties of $\phi(2170)$ observed in Ref.~\cite{Ablikim:2020pgw} seem to rule out the quark-antiquark or hybrid nature for $\phi(2170)$, while the tetraquark picture faces a challenge~\cite{Ding:2007pc,Dudek:2011bn}. In this work, we are going to consider the model of Ref.~\cite{MartinezTorres:2008gy}, in which $\phi(2170)$ is interpreted as a state generated from the dynamics involved in the $\phi K\bar K$ system, with $K\bar K$ resonating in the $f_0(980)$ region. To calculate the partial decay widths of $\phi(2170)$ to $K^+(1460) K^-$, $K^+_1(1400) K^-$ and $K^+_1(1270) K^-$, we also need a model to describe the properties of $K^+(1460)$, $K^+_1(1400)$ and $K^+_1(1270)$. In case of $K(1460)$, we follow Ref.~\cite{Torres:2011jt} and interpret $K(1460)$ as a $KK\bar K$ state with a large coupling to the $Kf_0(980)$ configuration  of the system, while for $K^+_1(1270)$ and $K^+_1(1400)$ we are going to adopt three different approaches: 
 (1) Treating $K^+_1(1270)$ and $K^+_1(1400)$ as a mixture of states belonging to the nonet of axial resonances. The experimental data on $K^+_1(1270)$ and $K^+_1(1400)$~\cite{Barbieri:1976mg,Carnegie:1977uz,Suzuki:1993yc,Blundell:1995au} are often analyzed by considering them as a mixture of two states~\cite{Barbieri:1976mg,Carnegie:1977uz,Suzuki:1993yc, Palomar:2003rb,Roca:2004uc}, typically named $K_{1A}$ and $K_{1B}$, which correspond to the strange partners of $a_1(1260)$ and $b_1(1235)$, respectively.  Although the exact value of the mixing angle is not well known, it could correspond to something between $\sim 20^\circ-45^\circ$~\cite{Barbieri:1976mg,Carnegie:1977uz,Suzuki:1993yc, Palomar:2003rb,Roca:2004uc}. 
 (2) Treating $K_1(1270)$ as a molecular state. In recent years, a double pole nature for $K_1(1270)$ has been claimed~\cite{Roca:2005nm,Geng:2006yb}. In these latter works, $K_1(1270)$ is interpreted as a superposition of two states generated from the unitarized dynamics of vector-pseudoscalar channels like $\phi K$, $\rho K$, $\pi K^*(892)$. Within the approach of Refs.~\cite{Roca:2005nm,Geng:2006yb}, $K^+_1(1400)$ does not appear, but we can determine the partial decay width of $\phi(2170)$ to each of the poles related to $K_1(1270)$ 
 (3) Alternatively to the previous two approaches, we can consider a phenomenological model based on the known data related to $K_1(1270)$ and $K_1(1400)$~\cite{PDG}.  Using such a model, we can determine the decay widths of $\phi(2170)\to K^+_1(1400)K^-,~K^+_1(1270) K^-$. 

As we will show, by considering $\phi(2170)$ as a $\phi K\bar K$ state, we obtain branching fractions which are compatible with those determined from the available $\mathcal{B}r\Gamma^{e^+e^-}_R$ results of Ref.~\cite{Ablikim:2020pgw}. 

\section{Formalism}\label{For}
\begin{figure}[h!]
\centering
\includegraphics[width=0.5\textwidth]{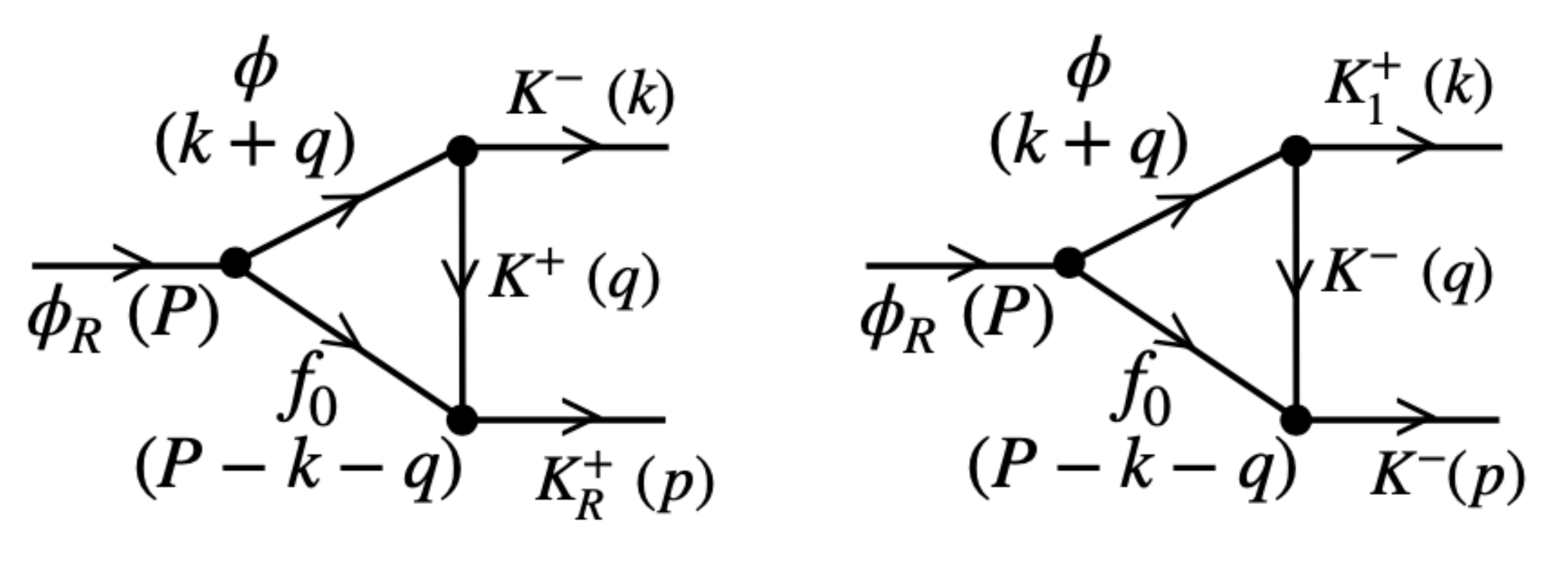}
\caption{Decay mechanism of $\phi(2170)$ to $K^+(1460) K^-$ (left), $K^+_1(1400)K^-$ and $K^+_1(1270) K^-$ (right). We use the nomenclature $\phi_R\equiv \phi(2170)$, $f_0\equiv f_0(980)$, $K_R\equiv K(1460)$ and $K_1$ can represent either $K_1(1400)$ or $K_1(1270)$.}\label{decay}
\end{figure}

We calculate the decay of $\phi(2170)$ to the channels $K^+(1460) K^-$, $K^+_1(1410) K^-$, and $K^+_1(1270) K^-$. To do this, we rely on the findings of Ref.~\cite{MartinezTorres:2008gy}, where $\phi(2170)$ is found to arise as a result of three-body interactions. In Ref.~\cite{MartinezTorres:2008gy} three-body scattering equations were solved for the $\phi K\bar K$ system, allowing each of the subsystems to interact in s-wave. As a consequence, a resonance was found to appear with mass around 2150 MeV when the $K\bar K$ subsystem interacts in isospin zero with an invariant mass $\sim980$ MeV. In other words, the $\phi(2170)$ resonance is found when the $\phi K\bar K$ system acts  effectively as  $\phi f_0(980)$. A study of a different three-body system, replacing $\phi$ by a kaon, was done in Ref.~\cite{Torres:2011jt}.
In this case, a  resonance with mass $\sim1460$ MeV was found when the $K\bar K$ system assembles itself as $f_0(980)$. The state obtained in Ref.~\cite{Torres:2011jt} was associated with $K(1460)$. Using the findings of Refs.~\cite{MartinezTorres:2008gy,Torres:2011jt} for $\phi(2170)$, $K(1460)$ and keeping in mind  that  $K_1(1270)$ and $K_1(1400)$ decay  to  vector-pseudoscalar channels with large branching ratios, we consider that $\phi(2170)$ decays to the aforementiond channels through the diagrams shown in Fig.~\ref{decay}. 

As can be seen, due to the nature and properties of the states involved, the processes $\phi(2170)\to K^+(1460) K^-,~K^+_1(1400) K^-,~K^+_1(1270) K^-$ proceed through a triangular loop of a virtual $\phi$, $f_0(980)$ and $K^\pm$  (henceforth, for the sake of convenience, we shall denote $\phi(2170)$ as $\phi_R$, $K(1460)$ as $K_R$, $f_0(980)$ as $f_0$ and use $K_1$ for $K_1(1400)$ and $K_1(1270)$ whenever there is no need to distinguish them). 

\begin{figure}[h!]
\centering
\includegraphics[width=0.7\textwidth]{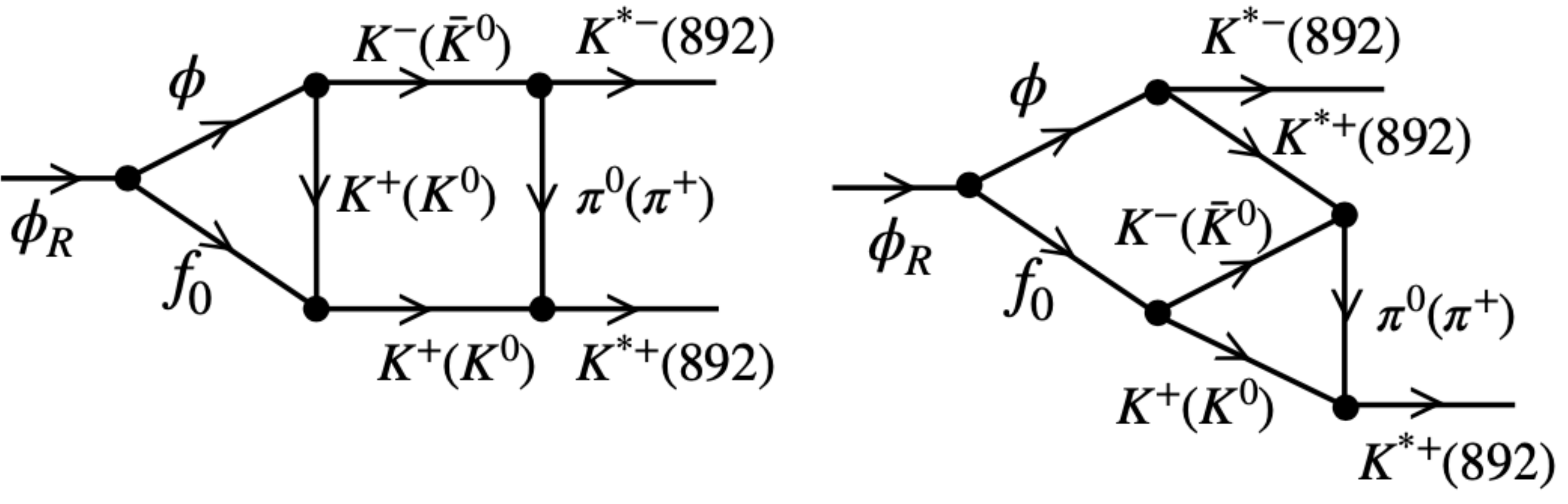}
\caption{Some of the decay mechanisms of $\phi(2170)$ to $K^{*+}(892) K^{*-}(892)$.}\label{decay2}
\end{figure}

Considering $\phi_R$ as a $\phi f_0(980)$ resonance, the situation is different for the decay process $\phi_R\to K^{*+}(892) K^{*-}(892)$ (see Fig.~\ref{decay2}). In this case the $\phi f_0(980)$ structure  of $\phi_R$ suppresses the decay to $K^{*+}(892) K^{*-}(892)$ as compared to the ones shown in Fig.~\ref{decay}. This is because the former process involves more than one loop (of triangular or higher topologies), as can be seen in Fig.~\ref{decay2}. Thus, within a $\phi f_0(980)$ molecular type description for $\phi(2170)$, one of the main conclusions of Ref.~\cite{Ablikim:2020pgw} gets naturally explained.

Let us now determine the amplitudes for the processes shown in Fig.~\ref{decay} to calculate the corresponding partial decay widths. To do this, we use the Lagrangian~\cite{Bando:1987br} 
\begin{align}
\mathcal{L}=-ig\langle V_\mu[P,\partial^\mu P]\rangle,\label{Lag}
\end{align}
to describe the $\phi K^+ K^-$ vertex, where $g=M_V/(2f_\pi)$ (with $M_V\simeq M_\rho$, $f_\pi\simeq 93$ MeV is the pion decay constant), $V_\mu$ and $P$ are matrices having as elements the vector and pseudoscalar meson fields, 
\begin{align}
P=\left(\begin{array}{ccc}\frac{\eta}{\sqrt{6}}+\frac{\pi^0}{\sqrt{2}}&\pi^+&K^+\\\pi^-&\frac{\eta}{\sqrt{6}}-\frac{\pi^0}{\sqrt{2}}&K^0\\K^-&\bar K^0&-\sqrt{\frac{2}{3}}\eta\end{array}\right),\quad V_\mu=\left(\begin{array}{ccc}\frac{\omega+\rho^0}{\sqrt{2}}&\rho^+&K^{*+}\\\rho^-&\frac{\omega-\rho^0}{\sqrt{2}}&K^{*0}\\K^{*-}&\bar K^{*0}&\phi\end{array}\right)_\mu.
\end{align}

The contribution of the vertices $\phi_R\phi f_0$, $K_R K f_0$, $K_1 \phi K$ and $f_0 K \bar K$ can be written in terms of the corresponding fields  as
\begin{align}
t_{\phi_R\to \phi f_0}&=g_{\phi_R\to \phi f_0}\epsilon_{\phi_R}\cdot \epsilon_\phi,\nonumber\\
t_{K^+_R\to K^+f_0}&=g_{K^+_R\to K^+ f_0},\nonumber\\
t_{f_0\to K^+ K^-}&=g_{f_0\to K^+ K^-},\nonumber\\
t_{K^+_1\to \phi K^+}&=g_{K^+_1\to \phi K^+}\epsilon_{K^+_1}\cdot \epsilon_\phi,\label{ts}
\end{align}
where $g_{\alpha\to \beta}$ represents the coupling of the state $\alpha=\phi_R,~K^+_R,~K^+_1,~f_0$ to the channel $\beta=\phi f_0,~K^+f_0,~\phi K^+,~K^+ K^-$, respectively. The coupling constants related to each vertex in Eq.~(\ref{ts}) depend on the properties of the hadrons involved in the vertex. In the following, we discuss the evaluation of these coupling constants.

\subsection{The $f_0(980) K^+ K^-$ vertex}\label{subsecf0}
There exists a growing evidence on the dominant role played by the $K\bar K$ dynamics in describing the properties of $f_0(980)$ (see the review on ``Interpretation of the scalars below 1 GeV'' of Ref.~\cite{PDG}). Based on the degrees of freedom of the different models, $f_0(980)$ is often described as a $s\bar s$ state surrounded by a $K\bar K$ meson cloud or as a $K\bar K$ bound state~\cite{PDG}. 

To determine the coupling $g_{f_0\to K^+ K^-}$, we follow the chiral unitary approach of Ref.~\cite{Oller:1997ti}, where $f_0$ is generated from the interaction of two pseudoscalars, in particular, $K\bar K$ and $\pi\pi$, in the isospin $I=0$ configuration. In this way, $g_{f_0\to (K \bar K)_{0}}$ (where the subscript indicates the isospin configuration of the system) can be calculated from the residue of the $T$-matrix in the complex energy plane, where a pole for $f_0(980)$ is found. The value obtained is 
\begin{align}
g_{f_0\to (K \bar K)_0}=3895+i 1328~\text{MeV}.\label{g980}
\end{align}
Using the isospin phase convention $|K^-\rangle=-\Big|I=\frac{1}{2},I_3=-\frac{1}{2}\Big\rangle$, the couplings $g_{f_0\to (K \bar K)_0}$ and $g_{f_0\to K^+ K^-}$ are related through a Clebsch-Gordan coefficient,
\begin{align}
g_{f_0\to K^+ K^-}=-\frac{g_{f_0\to (K \bar K)_0}}{\sqrt{2}}.
\end{align}
The coupling in Eq.~(\ref{g980}) leads to a branching fraction 
$\Gamma\left(f_0\to\pi\pi\right)/[\Gamma\left(f_0\to\pi\pi )+ \Gamma(f_0\to K \bar K\right)]$~\cite{Oller:1997ti} compatible with the values known from experimental data~\cite{PDG}.

\subsection{The $\phi(2170)\phi f_0(980)$ vertex}
Motivated by the findings of our previous work~\cite{MartinezTorres:2008gy}, we describe $\phi(2170)$ as an effective  $\phi f_0(980)$ state.
In Ref.~\cite{MartinezTorres:2008gy}, $\phi(2170)$ is generated from the $\phi K\bar K$ interaction, with $K\bar K$ forming $f_0(980)$ in the energy region of the three-body resonance. The coupling of $\phi(2170)$ to $\phi f_0(980)$ can be determined in the following way~\cite{MartinezTorres:2008gy}: we assume that around the peak position, the scattering matrix  $T_{\phi f_0\to \phi f_0}$, which depends on the invariant mass of the $\phi f_0(980)$ system ($\sqrt{s_{\phi f_0}}$), is proportional to the three-body amplitude $T_{\phi (K\bar K)_0\to \phi (K\bar K)_0}$ when $\sqrt{s}\sim M_{\phi(2170)}$ and $\sqrt{s_{K\bar K}}\sim M_{f_0(980)}$, i.e.,
\begin{align}
T_{\phi f_0\to \phi f_0}=\alpha\, T_{\phi (K\bar K)_{0}\to \phi (K\bar K)_{0}}.\label{T}
\end{align}
In Eq.~(\ref{T}), $\alpha$ is a constant determined by imposing the unitary condition for $\text{Im}\{T^{-1}_{\phi f_0\to \phi f_0}\}$, treating the $\phi f_0(980)$ system as an effective two-body system, i.e.,
\begin{align}
\text{Im}\{T^{-1}_{\phi f_0\to \phi f_0}\}=\frac{|\vec{p}_{\phi f_0}|}{8\pi\sqrt{s_{\phi f_0}}},\label{Im}
\end{align}
with $|\vec{p}_{\phi f_0}|$ being the modulus of the center of mass momentum for the $\phi f_0$ system at $\sqrt{s_{\phi f_0}}\sim M_{\phi_R}$. If we assume now a Breit-Wigner form for $T_{\phi (K\bar K)_0\to \phi (K\bar K)_0}$ around $\sqrt{s}\sim M_{\phi_R}$ and $\sqrt{s_{K\bar K}}\sim M_{f_0}$, we can obtain the coupling $g_{\phi_R\to \phi f_0}$ in terms of the three-body amplitude given in Ref.~\cite{MartinezTorres:2008gy} as
\begin{align}
g^2_{\phi_R\to \phi(K\bar K)_0}=i M_{\phi_R}\Gamma_{\phi_R}T_{\phi (K\bar K)_0\to \phi (K\bar K)_0}.\label{g3B}
\end{align}
In this way, by using Eq.~(\ref{T}), we can get the coupling of $\phi(2170)$ to $\phi f_0(980)$ as
\begin{align}
g^2_{\phi_R\to \phi f_0}=\alpha g^2_{\phi_R\to \phi(K\bar K)_0}. \label{grel}
\end{align}
In the model of Ref.~\cite{MartinezTorres:2008gy}, the partial decay width of $\phi(2170)\to \phi (K\bar K)_0$ was found to be of the order of $30$ MeV. However, since the relation in Eq.~(\ref{T}) is meaningful for $\sqrt{s}\sim M_{\phi_R}$ and $\sqrt{s_{K\bar K}}\sim M_{f_0}$, we should admit certain uncertainty in the partial decay width of $\phi(2170)\to \phi f_0(980)$ and, thus, in the coupling $g_{\phi_R\to \phi f_0}$. To do this, we have considered that the partial decay width of $\phi(2170)\to \phi f_0(980)$ could change in the range $30-50$ MeV, while keeping the strength of $T_{\phi (K\bar K)_0\to \phi (K\bar K)_0}$ around the peak position. We then determine the average value and the standard deviation of $g_{\phi_R\to\phi f_0}$ when changing $\Gamma_{\phi_R\to \phi f_0}\sim 30-50$ MeV, and find
\begin{align}
|g_{\phi_R\to \phi f_0}|=3123\pm561~\text{MeV}.\label{gR}
\end{align}
Since the partial decay widths of $\phi(2170)$ depend, among other variables, on $|g_{\phi_R\to\phi f_0}|$, it is important to show the reliability of the value in Eq.~(\ref{gR}). To do this, we evaluate the cross sections for the $\phi f_0(980)$ configuration of the final state $K^+ K^- \pi^+(\pi^0) \pi^- (\pi^0)$, i.e., for the process $e^+ e^-\to \phi f_0(980)$, which is precisely the reaction in which $\phi(2170)$ was observed for the first time. The cross sections for $e^+ e^-\to \phi f_0(980)$ have been determined by the Babar Collaboration in Refs.~\cite{Aubert:2006bu,Lees:2011zi}.

\begin{figure}[h!]
\includegraphics[width=0.58\textwidth]{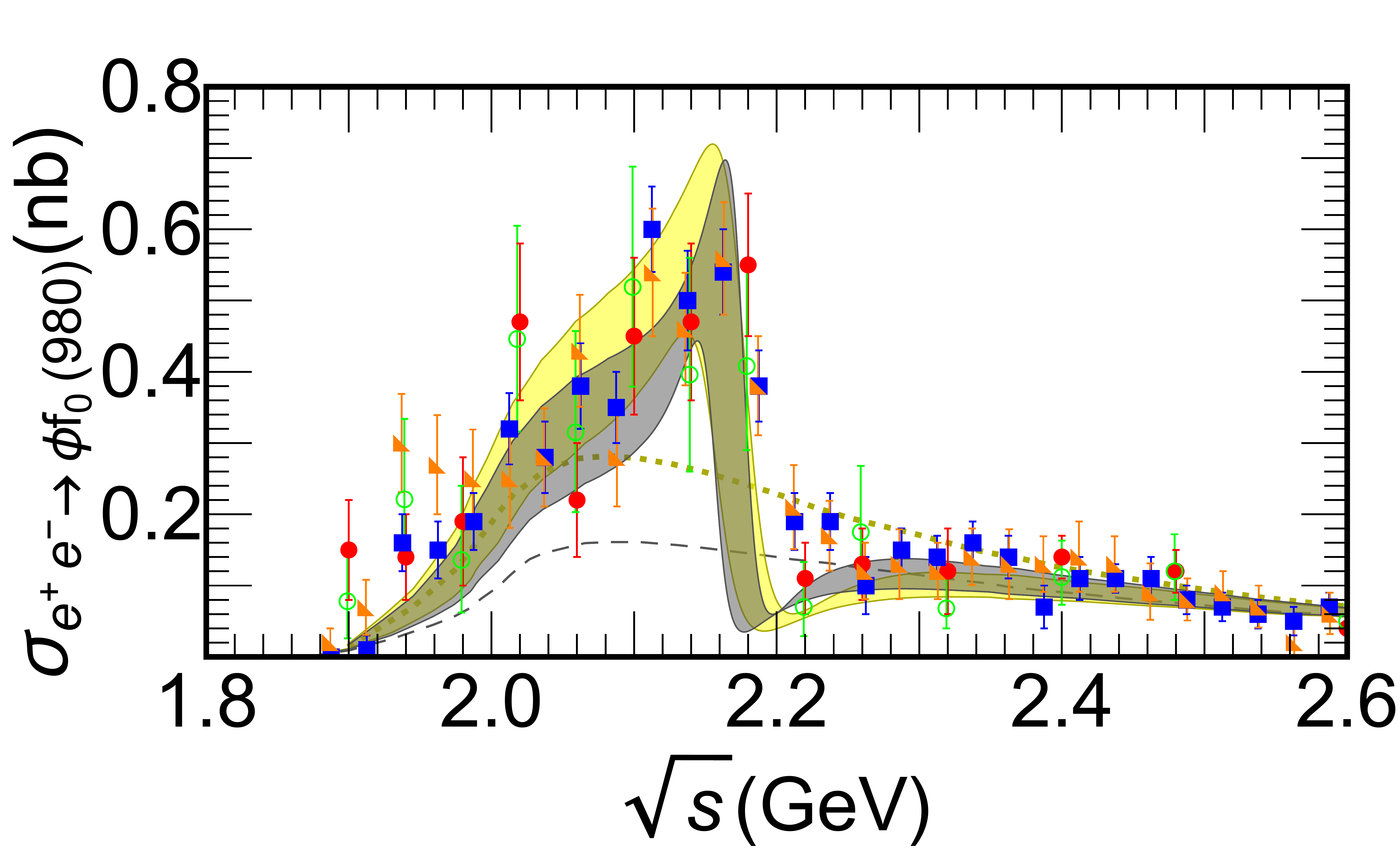}
\caption{Model results for the cross sections $e^+e^-\to \phi f_0(980)$ (data points are taken from Refs.~\cite{Aubert:2006bu,Lees:2011zi}). The dashed (dotted) line represents the background used in Ref.~\cite{MartinezTorres:2008gy} (Ref.~\cite{Lees:2011zi}). The lower and upper limits of  the shaded regions are obtained by using the above mentioned backgrounds when implementing the final state interaction to generate $\phi(2170)$.  The partial decay width of $\phi_R\to \phi f_0$ is changed between 30 MeV (dark shaded region) to 50 MeV (light shaded region).}\label{comp}
\end{figure}

In Fig.~\ref{comp} we show the results found for the $e^+e^-\to \phi f_0(980)$ cross sections. The different data sets correspond to the Babar data for the cross sections of $e^+ e^-\to \phi f_0(980)$ collected over different years (empty circles and triangles are taken from Ref.~\cite{Aubert:2006bu}; filled circles and squares are from Ref.~\cite{Lees:2011zi}). The dark (light) shaded region represents the cross sections obtained within our model (described in appendix~\ref{AphiR}) by considering a partial decay width of $\phi(2170)\to \phi f_0(980)$ of $\sim$30 (50) MeV.  The lower (upper) bound of these regions represents the result obtained with the background found in Ref.~\cite{Napsuciale:2007wp} (Ref.~\cite{Lees:2011zi}) for the process $e^+e^-\to \phi f_0(980)$ and a peak position for $\phi(2170)$ of 2150 (2175) MeV, as in Ref.~\cite{MartinezTorres:2008gy} (Ref.~\cite{Lees:2011zi}). 
As can be seen in Fig.~\ref{comp}, the data on $e^+ e^-\to \phi f_0$ are well reproduced, which shows the suitability of the interpretation of $\phi(2170)$ as a $\phi f_0(980)$ state and the value obtained for $g_{\phi_R\to \phi f_0}$.

\subsection{The $K^+(1460) K^+ f_0(980)$ vertex}
To get the coupling $g_{K^+_R\to K^+ f_0}$, we rely on the findings of Ref.~\cite{Torres:2011jt}, where three-kaon scattering equations were solved within two different formalisms. One of the methods in Ref.~\cite{Torres:2011jt} consisted of solving Faddeev equations with unitarized chiral two-body amplitudes for $KK\bar K$, $K\pi\pi$ and $K\pi\eta$ coupled systems. All the two-body interactions were kept in s-wave. Within a second method, a nonrelativistic potential model was used to study the three-kaon system to obtain the corresponding wavefunction through the variational approach.  In both cases, a three-body resonance was found with mass in the range of 1420-1460 MeV and width varying between 50-100 MeV, when one of the $K\bar K$ system forms $f_0(980)$.  The state was related to $K(1460)$. The $KK\bar K$ s-wave interactions have been studied within several approaches different to the one used in Ref.~\cite{Torres:2011jt} (see Refs.~\cite{Albaladejo:2010tj,Kezerashvili:2015aca,Shinmura:2019nqw,Filikhin:2020ksv}), and a kaon state has always been found to arise with mass $\sim1460$ MeV but with widths ranging between 50-200 MeV, depending on the model. 

With the findings of Ref.~\cite{Torres:2011jt} at hand, one would imagine that analogously to the case of $\phi(2170)\phi f_0(980)$, the coupling $g_{K^+_R\to K^+ f_0}$ can be obtained by relating the $T$-matrices $T_{K(K\bar K)_0\to K(K\bar K)_0}$ and $T_{K f_0\to K f_0}$  via Eqs.~(\ref{T}) and (\ref{Im}). However,  $K(1460)$ is below the $Kf_0(980)$ threshold, contrary to $\phi(2170)$, and, thus, relations based on the unitary condition cannot be used. 
\begin{figure}[h!]
\centering
\includegraphics[width=0.3\textwidth]{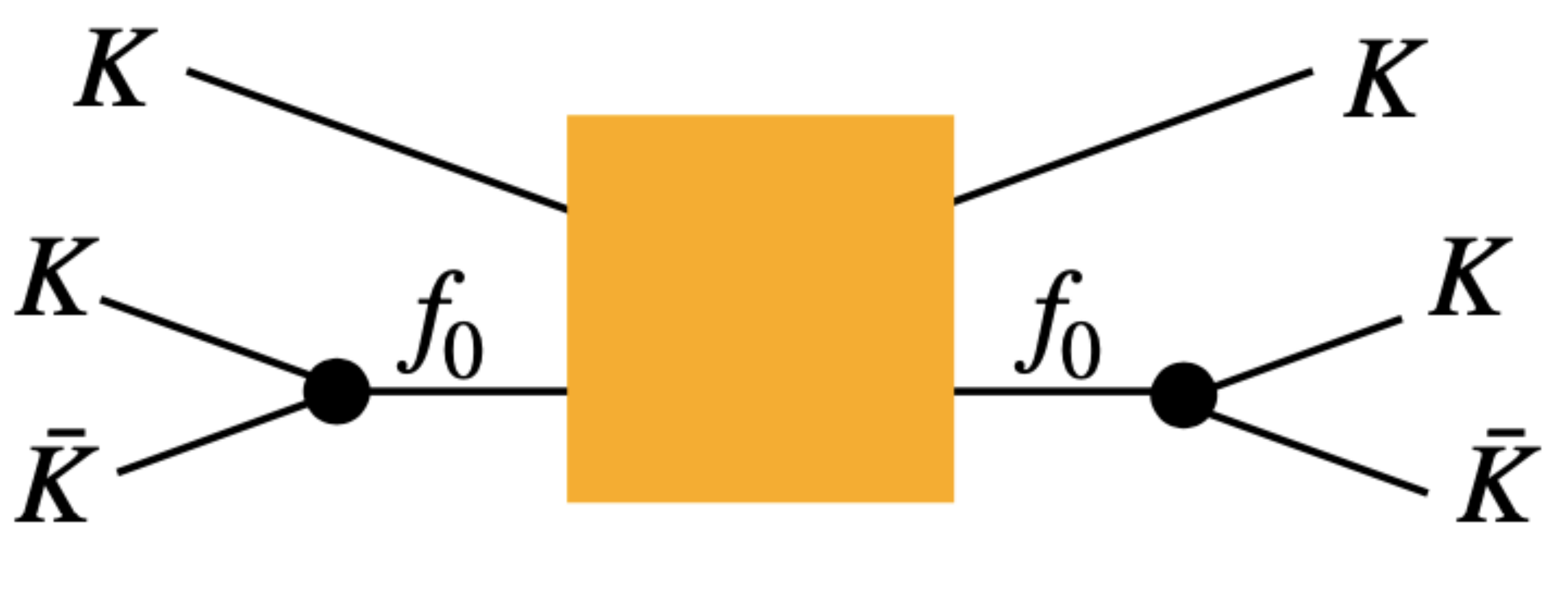}
\caption{Formation of $K(1460)$ when a $K$ and a $\bar K$ in the isospin 0 configuration interact to generate the $f_0(980)$ resonance.}\label{KR}
\end{figure}
A different strategy to calculate $g_{K^+_R\to K^+ f_0}$ is to consider that since the $T$-matrix for the $K(K\bar K)_0$ system depends on $\sqrt{s}$ and $\sqrt{s_{(K\bar K)_0}}$, $T_{K(K\bar K)_0\to K(K\bar K)_0}$ and $T_{K f_0\to K f_0}$ can be related via~\cite{MartinezTorres:2011vh} (based on the diagram shown in Fig.~\ref{KR})
\begin{align}
T_{K(K\bar K)_0\to K(K\bar K)_0}=\left[\frac{g_{f_0\to (K\bar K)_0}}{s_{K\bar K}-M^2_{f_0}+i M_{f_0}\Gamma_{f_0}}\right]^2 T_{K f_0\to K f_0},\label{Tkf01}
\end{align}
for $\sqrt{s}\sim M_{K_R}$ and $\sqrt{s_{(K\bar K)_0}}\sim M_{f_0}$. By considering a Breit-Wigner form for $T_{K f_0\to K f_0}$ around $\sqrt{s_{Kf_0}}\sim M_{K_R}$, 
 \begin{align}
 T_{K f_0\to K f_0}=\frac{g^2_{K_R\to K f_0}}{s_{K f_0}-M^2_{K_R}+i M_{K_R}\Gamma_{K_R\to K f_0}},\label{Tkf02}
 \end{align}
we can determine the value of $g_{K_R\to K f_0}$ from Eq.~(\ref{Tkf01}). Interestingly, this way of finding $g_{K_R\to K f_0}$, if applied to the case of $\phi(2170)$ and the $\phi f_0$ system,  results in a value of $|g_{\phi_R\to \phi f_0}|$ similar to the one given in Eq.~(\ref{gR}).

Keeping in mind that the peak position and width of $K(1460)$, in Ref.~\cite{Torres:2011jt}, varies between 1420-1460 MeV and 50-110 MeV, respectively,  depending on the model used, we compute the corresponding uncertainties in the value of $g_{K_R\to K f_0}$. This is done by using a Breit-Wigner  description ($T^{BW}_{K(K\bar K)_0}$) for the three-body amplitude $T_{K(K\bar K)_0\to K(K\bar K)_0}$ obtained in Ref.~\cite{Torres:2011jt}, considering that  $|T^{BW}_{K(K\bar K)_0}| \simeq|T_{K(K\bar K)_0\to K(K\bar K)_0}|$ around the peak position, while varying the width in the range 50-110 MeV.  Using the mass and width  for $f_0(980)$ within the model~\cite{Oller:1997ti} discussed in section~\ref{subsecf0}, we get the following average and  standard deviation for  $|g_{K_R\to K f_0}|$ 
\begin{align}
|g_{K^+_R\to K^+ f_0}|=4858\pm 1337~\text{MeV}.\label{gKR}
\end{align}

To end this subsection, a brief discussion on the width of the state obtained in Ref.~\cite{Torres:2011jt} is in order here.  We determine Eq.~(\ref{gKR}) considering that the state found in Ref.~\cite{Torres:2011jt} corresponds to  $K(1460)$~\cite{PDG}, though the width of the state in Ref.~\cite{Torres:2011jt} is smaller than the one listed in Ref.~\cite{PDG} for $K(1460)$, $\sim 250-330$ MeV.  A possible reason for such an apparent discrepancy could be the fact that the width obtained in Ref.~\cite{Torres:2011jt} comes from three-body channels, considering $s$-wave interactions between the different pairs. According to Ref.~\cite{PDG}, $K(1460)$ decays to $K\pi\pi$ in $s$-wave as well as to $\pi K^*(892)$ in $p$-wave. Such $p$-wave channels can increase the decay width of $K(1460)$, even though they are not essential for the generation of $K(1460)$ in the approach of Ref.~\cite{Torres:2011jt}. From the study of Ref.~\cite{Aaij:2017kbo}, the partial decay widths of $K(1460)$ to $K\pi\pi$ in $s$-wave and to $\pi K^*(892)$ in $p$-wave are similar. Thus, the partial decay width of $K(1460)$ to $K\pi\pi$ in $s$-wave could be of  the order of 100 MeV, in line with the findings of  Ref.~\cite{Torres:2011jt}. And it is the $Kf_0$ coupling to $K(1460)$ which is relevant for the study of the $\phi(2170)$ decay, with the latter being interpreted as $\phi f_0$ state. Alternatively, it may be that the width of $K(1460)$ is overestimated in the partial wave analyses when fitting the data. 
For example, an interesting feature can be noticed in Figs.~1 and 3 of the supplemental material of Ref.~\cite{Ablikim:2020pgw}, which shows the data for the $e^+ e^-\to K^+ K^- \pi^0\pi^0$ reaction, for center of mass energies 2125 MeV and 2396 MeV. It can be seen that the energy dependence of the $K^\pm \pi^0\pi^0$ invariant mass distribution passes from having a broad distribution around 1460 MeV to a much richer structure, as the total energy increases. Further, though the $K^\pm\pi^0$ invariant mass distribution shows the formation of $K^*(892)$ at both the center of mass energies, a clear signal of $f_0(980)$ in the $\pi^0\pi^0$ invariant mass distribution is observed only at the center of mass energy of 2396 MeV. And it is at this center of mass energy, where, within the uncertainties in the $K^\pm \pi^0\pi^0$ data, a clearer structure at 1460 MeV seems to appear,  which could have a width narrower than the extracted value of $230\pm 35$ MeV. This indicates that experimental data with higher statistics can be useful to clarify the properties of $K(1460)$. 

\subsection{The $K^+_1\phi K^+$ vertex}
The nature of $K_1(1270)$ and $K_1(1400)$ is still under debate. One of the approaches frequently used to describe the properties of these states is to consider that they are both a mixture of the $K_{1A}$ and $K_{1B}$ states, belonging to the nonet of axial resonances. In the last decades,
a different nature has been proposed for the axial resonances~\cite{Lutz:2003fm,Roca:2005nm,Geng:2006yb,Nagahiro:2008zza,Geng:2008ag,Zhou:2014ila}. In these models, axial resonances are found to arise from the interaction of pseudoscalar and vector mesons. Alternatively to the above two approaches, we can consider a phenomenological model using the known data related to $K_1(1270)$ and $K_1(1400)$~\cite{PDG} to study the decay of $\phi(2170)$.

Thus, in this work, to evaluate the couplings $g_{K^+_1(1270)\to \phi K^+}$ and $g_{K^+_1(1400)\to \phi K^+}$ we have considered the aforementioned three different approaches. We discuss more details on the determination of $g_{K^+_1(1270)\to \phi K^+}$ and $g_{K^+_1(1400)\to \phi K^+}$  in the following subsections.

\subsubsection{Model A: $K_1(1270)$ as a state arising from meson-meson dynamics} \label{double}
In Refs.~\cite{Lutz:2003fm,Roca:2005nm,Geng:2006yb,Nagahiro:2008zza,Geng:2008ag,Zhou:2014ila} (mainly in Refs.~\cite{Lutz:2003fm,Roca:2005nm}, with the other references being works following the latter one) two poles have been found around 1270 MeV with the quantum numbers of $K_1(1270)$. However, the interpretation of the poles is different in Refs.~\cite{Lutz:2003fm,Roca:2005nm}. While the former work associates the two poles with $K_1(1270)$ and $K_1(1400)$, both poles have been related to $K_1(1270)$ in the latter one and it is argued that the superposition of the two poles should be interpreted as the signal observed in the experimental data for $K_1(1270)$. Further, it has been shown in Ref.~\cite{Geng:2006yb} that the double pole nature for $K_1(1270)$ describes well the WA3 data on $K^- p\to K^-\pi^+\pi^- p$. To consider the influence of the molecular nature of $K_1(1270)$ on the decay of $\phi(2170)$ to $ K_1(1270) \bar K$, we use the information given in Ref.~\cite{Geng:2006yb} on the $K_1(1270) \phi K$ couplings, since it is straightforward to implement in our model describing $\phi(2170)$ as a $\phi f_0$ resonance.
For the convenience of the reader, we list here the poles, as found in Ref.~\cite{Geng:2006yb}, 
\begin{align}
z_1&=1195-i 123~\text{MeV},\nonumber\\
z_2&=1284-i 73~\text{MeV}, \label{z12}
\end{align}
and their couplings to the $\phi K$ channel
\begin{align}
 g^{(1)}_{K_1(1270)\to \phi K}&=2096-i1208~\text{MeV},\nonumber\\
 g^{(2)}_{K_1(1270)\to \phi K}&=1166-i774~\text{MeV},\label{g12}
 \end{align}
with the superscript being related to the poles in Eq.~(\ref{z12}). When calculating the decay width of $\phi(2170)\to K^+_1(1270) K^-$ we have considered the contribution of each pole separately as well as the superposition of the two poles.

Since the state $K_1(1400)$ is not generated within the approach of Ref.~\cite{Geng:2006yb}, the decay width of $\phi(2170)\to K^+_1(1400)K^-$ is not determined within such an interpretation.
  
\subsubsection{Model B: $K_1(1270)$ and $K_1(1400)$ within a mixing scheme}\label{mixed}

To consider $K_1(1270)$ and $K_1(1400)$ as a mixture of $K_{1A}$ and $K_{1B}$, states belonging to the nonet of axials, we use the information given in Ref.~\cite{Palomar:2003rb} on the couplings of $K_1(1270)$ and $K_1(1400)$ to different pseudoscalar-vector channels. Such couplings (in Table 3 of Ref.~\cite{Palomar:2003rb} ) lead to partial decay widths of axial resonances to hadronic channels which are compatible with the values known from experimental data.  
It must be mentioned that the mixing angle between $K_{1A}$ and $K_{1B}$ is not known with precision and different values have been determined phenomenologically. We will use values of $29^\circ$, $47^\circ$ and $62^\circ$, which have been claimed to be compatible with the data~\cite{Barbieri:1976mg,Carnegie:1977uz,Suzuki:1993yc, Palomar:2003rb,Roca:2004uc}.

The couplings provided in Ref.~\cite{Palomar:2003rb} cannot be directly used here, since they are related to vector mesons described by tensor fields of rank 2~\cite{Ecker:1988te,Xiong:1992ui}. The vector meson field in our work [see the $t_{K^+_1\to \phi K^+}$ amplitude in Eq.~(\ref{ts})], on the other hand, is written in terms of an associated polarization vector~\cite{Bando:1984ej,Bando:1987br}. Thus, to determine the coupling $g_{K^+_1\to \phi K^+}$ which should be used in Eq.~(\ref{ts}) we first evaluate the decay width of $K_1\to \phi K$ within the approach of Ref.~\cite{Palomar:2003rb}, $\Gamma^T_{K^+_1\to \phi K^+}$, and obtain the value of $g_{K^+_1\to \phi K^+}$ such as to reproduce the same width with Eq.~(\ref{ts}) (see appendix~\ref{widthK1} for the details on the calculation of the decay width).

We find the following values of $|g_{K^+_1\to \phi K^+}|$ as a function of the mixing angle $\alpha$ 
\begin{align}
|g_{K^+_1(1270)\to \phi K^+}|=\left\{\begin{array}{c}1081~\text{MeV},~\alpha=29^\circ,\\ 946~\text{MeV},~\alpha=47^\circ,\\1286~\text{MeV},~\alpha=62^\circ,\end{array}\right.\quad
|g_{K^+_1(1400)\to \phi K^+}|=\left\{\begin{array}{c}3543~\text{MeV},~\alpha=29^\circ,\\ 3546~\text{MeV},~\alpha=47^\circ,\\3509~\text{MeV},~\alpha=62^\circ,\end{array}\right.\label{gmixed}
\end{align}
To take into account the uncertainty in the value of the mixing angle, we calculate the average and standard deviation for the values of the couplings  using the above results, and we get
\begin{align}
|g_{K^+_1(1270)\to \phi K^+}|&= 1104\pm171~\text{MeV},\nonumber\\
|g_{K^+_1(1400)\to \phi K^+}|&=3533\pm21~\text{MeV}.
\end{align}

  
\subsubsection{Model C: A phenomenological approach to describe the $K_1\phi K$ vertex}\label{pheno}
Instead of considering a molecular nature for $K_1(1270)$ or using an approach based on treating $K_1(1270)$ and $K_1(1400)$ as mixture of states belonging to axial nonets, we can determine the couplings of $K_1(1270)$ and $K_1(1400)$ to $\phi K$ phenomenologically, by using the available data on the hadronic and radiative decay of these states~\cite{PDG}. We refer the reader to appendix~\ref{radiative} for the details on the evaluation of the coupling $K_1\to\phi K$ within this model. 

Given the uncertainties in the experimental data, three different solutions are found for $|g_{K^+_1(1270)\to \phi K^+}|$, which are
\begin{align}
|g_{K^+_1(1270)\to \phi K^+}|=\left\{\begin{array}{c}3967\pm 419~\text{MeV},~\text{Solution $\mathbb{S}_1$},\\~12577\pm763~\text{MeV},~\text{Solution $\mathbb{S}_2$},\\~19841\pm1177~\text{MeV},~\text{Solution $\mathbb{S}_3$}.\end{array}\right.\label{gK1phen}
\end{align}
In case of the process $K^+_1(1400)\to \phi K^+$ we obtain the following value 
\begin{align}
|g_{K^+_1(1400)\to \phi K^+}|=8480\pm 1333~\text{MeV}.\label{gK114}
\end{align}

A word of caution is here in order: it is important to recall that information from direct measurements of processes like $K_1(1270)\to \gamma K$ is not available and the radiative decay widths of $K_1(1270)$ and $K_1(1400)$ are extracted through Primakoff effect, by assuming that they are a mixture of the $K_{1A}$ and $K_{1B}$ states mentioned in the previous section. Thus, if the $K_1$ resonances have a different origin, and are not related through a mixing angle, then the experimental information available~\cite{PDG} on the radiative decay widths of $K_1$, and, hence, the results obtained on $\phi(2170)\to K^+_1(1270) K^-$ and $K^+_1(1400) K^-$ within this phenomenological approach, will be required to be revised.

\subsection{Decay widths of $\phi(2170)$ into a kaonic resonance plus a $\bar K$}\label{sw}
The decay widths for the processes shown in Fig.~\ref{decay} can be obtained as
\begin{align}
\Gamma=\int d\Omega \frac{|\vec{p}|}{32\pi^2 M^2_{\phi(2170)}}\overline{\sum\limits_\text{pol}}|t|^2,\label{wphiR}
\end{align}
where $\int d\Omega$ is the solid angle integration, $t$ represents the amplitude for each of the processes depicted in Fig.~\ref{decay}, and the symbol $\overline{\sum\limits_{\text{pol}}}$ indicates sum over the polarizations of the particles in the initial and final states, and average over the polarizations of the particles in the initial state.

Using the Feynman rules, we can write the amplitudes necessary to calculate Eq.~(\ref{wphiR}) in terms of the vertices described in the previous section. In case of the process $\phi(2170)\to K^+(1460) K^-$, we have 
\begin{align}
-it_{\phi_R\to K^+_R K^-}&=\int\frac{d^4q}{(2\pi)^4}t_{\phi_R\to\phi f_0} t_{\phi\to K^+ K^-} t_{f^0 K^+\to K^+(1460)}\nonumber\\
&\quad\times\frac{1}{[(k+q)^2-M^2_\phi+i\epsilon][(P-k-q)^2-M^2_{f_0}+i\epsilon][q^2-M^2_K+i\epsilon]}.
\end{align}
Considering now Eqs.~(\ref{Lag}) and~(\ref{ts}), we get after summing over the polarizations of the internal vector mesons,
\begin{align}
-it_{\phi_R\to K^+_R K^-}=&g_{\phi_R\to\phi f_0}gg_{K^+_R\to K^+f_0}\epsilon^\mu_{\phi_R}(P)\nonumber\\
&\quad\times\Bigg[k_\mu\left(1-\frac{k^2}{M^2_\phi}\right)I^{(0)}-I^{(1)}_\mu\left(1+\frac{k^2}{M^2_\phi}\right)+\frac{k_\mu}{M^2_\phi}I^{(2)}+\frac{I^{(3)}_\mu}{M^2_\phi}\Bigg],\label{tKR}
\end{align}
where
\begin{align}
I^{(0)};I^{(1)}_\mu;I^{(2)}_{\mu\nu};I^{(2)}&\equiv\int\frac{d^4q}{(2\pi)^4}\frac{1;q_\mu;q_\mu q_\nu;q^2}{\mathcal{D}},\label{Int}
\end{align}
and 
\begin{align}
\mathcal{D}=[(k+q)^2-M^2_\phi+i\epsilon][(P-k-q)^2-M^2_{f_0}+i\epsilon][q^2-M^2_K+i\epsilon].
\end{align}
Similarly, for $\phi(2170)\to K^+_1 K^-$, 
\begin{align}
-it_{\phi_R\to K^+_1 K^-}&=g_{\phi_R\to\phi f_0}g_{K^+_1\to\phi K^+}g_{f_0\to K^+ K^-}\epsilon^\mu_{\phi_R}(P)\epsilon^\nu_{K^+_1}(k)\nonumber\\
&\quad\times\Bigg[-g_{\mu\nu}I^{(0)}+\frac{k_\mu}{M^2_\phi}I^{(1)}_\nu+\frac{I^{(2)}_{\mu\nu}}{M^2_\phi}\Bigg],\label{tK1}
\end{align}
where $K^+_1$ can be $K^+_1(1270)$ or $K^+_1(1400)$. 

Next, we need to calculate the expressions in Eq.~(\ref{Int}). To do this, we consider the 
Passarino-Veltman reduction for tensor integrals~\cite{Passarino:1978jh} and write
\begin{align}
I^{(1)}_\mu&=a^{(1)}_1k_\mu+a^{(1)}_2 P_\mu,\nonumber\\
I^{(3)}_\mu&=a^{(3)}_1k_\mu+a^{(3)}_2 P_\mu,\nonumber\\
I^{(2)}_{\mu\nu}&=a^{(2)}_1g_{\mu\nu}+a^{(2)}_2(k_\mu P_\nu+k_\nu P_\mu)+a^{(2)}_3k_\mu k_\nu+a^{(2)}_4P_\mu P_\nu,\label{Icov}
\end{align}
where $a^{(i)}_j$ are coefficients to be calculated. In this way, we can write Eqs.~(\ref{tKR}) and (\ref{tK1}) as
\begin{align}
-it_{\phi_R\to K^+_R K^-}&=g_{\phi_R\to\phi f_0}gg_{K^+_R\to K^+ f_0}\epsilon_{\phi_R}(P)\cdot k\nonumber\\
&\quad\times\Bigg[-a^{(1)}_1+I^{(0)}+\frac{1}{M^2_\phi}\left(a^{(3)}_1+I^{(2)}-k^2\left\{a^{(1)}_1+I^{(0)}\right\}\right)\Bigg],\nonumber\\
-it_{\phi_R\to K^+_1 K^-}&=g_{\phi_R\to\phi f_0}g_{K^+_1\to \phi K^+}g_{f_0\to K^+ K^-}\epsilon^\mu_{\phi_R}(P)\epsilon^\nu_{K^+_1}(k)\nonumber\\
&\quad\times\Bigg[g_{\mu\nu}\left(-I^{(0)}+\frac{a^{(2)}_1}{M^2_\phi}\right)+\frac{k_\mu P_\nu}{M^2_\phi}\left(a^{(1)}_2+a^{(2)}_2\right)\Bigg], 
\end{align}
where we have used the Lorenz gauge condition. Thus, to get the amplitudes written above and the corresponding decay widths, we need to determine the coefficients $a^{(1)}_1$, $a^{(1)}_2$, $a^{(2)}_1$, $a^{(2)}_2$, $a^{(3)}_1$ and the integrals $I^{(0)}$ and $I^{(2)}$. To do this, we proceed as follows: to calculate the coefficients $a^{(1)}_1$, $a^{(1)}_2$, we contract the tensor $I^{(1)}_\mu$ in Eq.~(\ref{Icov}) with the different Lorentz structures present there, forming a system of coupled equations, i.e.,
\begin{align}
k\cdot I^{(1)}&=a^{(1)}_1k^2+a^{(1)}_2 k\cdot P,\nonumber\\
P\cdot I^{(1)}&=a^{(1)}_1k\cdot P+a^{(1)}_2 P^2.
\end{align}
In this way, the coefficients $a^{(1)}_1$ and $a^{(1)}_2$ can be written in terms of the scalar integrals $k\cdot I^{(1)}$ and $P\cdot I^{(1)}$ as
\begin{align}
a^{(1)}_1&=-\frac{P^2(k\cdot I^{(1)})-k\cdot P (P\cdot I^{(1)})}{(k\cdot P)^2-k^2 P^2},\nonumber\\
a^{(1)}_2&=-\frac{k^2(P\cdot I^{(1)})-k\cdot P (k\cdot I^{(1)})}{(k\cdot P)^2-k^2 P^2}.\label{a1}
\end{align}

Similarly, considering now the tensors $I^{(3)}_\mu$ and $I^{(2)}_{\mu\nu}$ in Eq.~(\ref{Icov}) and following the same procedure, we arrive to
\begin{align}
a^{(3)}_1&=-\frac{P^2(k\cdot I^{(3)})-k\cdot P (P\cdot I^{(3)})}{(k\cdot P)^2-k^2 P^2},\nonumber\\
a^{(3)}_2&=-\frac{k^2(P\cdot I^{(3)})-k\cdot P (k\cdot I^{(3)})}{(k\cdot P)^2-k^2 P^2}.\label{a3}
\end{align}
and
\begin{align}
a^{(2)}_1&=\frac{1}{2[(k\cdot P)^2-k^2 P^2]}\Big[k^2\left(P\cdot P\cdot I^{(2)}-I^{(2)} P^2\right)+I^{(2)}(k\cdot P)^2\nonumber\\
&\quad-2(k\cdot P\cdot I^{(2)})(k\cdot P)+(k\cdot k\cdot I^{(2)})P^2\Big],\nonumber\\
a^{(2)}_2&=\frac{1}{2[(k\cdot P)^2-k^2 P^2]^2}\Big[k^2\left\{(I^{(2)}P^2-3 P\cdot P\cdot I^{(2)})(k\cdot P)+2(k\cdot P\cdot I^{(2)})P^2\right\}\nonumber\\
&\quad-(k\cdot P)\left\{I^{(2)}(k\cdot P)^2-4(k\cdot P\cdot I^{(2)})(k\cdot P)+3(k\cdot k\cdot I^{(2)})P^2\right\}\Big].\label{a2}
\end{align}
Going back to Eq.~(\ref{Int}) and integrating on the $q^0$ variable using Cauchy's theorem, we can obtain the scalar integrals appearing in Eqs.~(\ref{tKR}),~(\ref{tK1}), (\ref{a1}), (\ref{a3}) and (\ref{a2}). Considering the rest frame of the decaying particle, i.e., $\vec{P}=\vec{0}$, we get
\begin{align}
&I^{(0)}=\int\frac{d^3q}{(2\pi)^3}\mathbb{I}_0,\quad k\cdot I^{(1)}=\int\frac{d^3q}{(2\pi)^3}[k^0\mathbb{I}_1-\vec{k}\cdot\vec{q}\,\mathbb{I}_0],\quad P\cdot I^{(1)}=\sqrt{s}\int\frac{d^3q}{(2\pi)^3}\mathbb{I}_1,\nonumber\\
&I^{(2)}=\int \frac{d^3q}{(2\pi)^3}(\mathbb{I}_2-\vec{q}^{\,2}\mathbb{I}_0),\quad k\cdot P\cdot I^{(2)}=\sqrt{s}\int \frac{d^3q}{(2\pi)^3}[k^0 \mathbb{I}_2-(\vec{k}\cdot\vec{q} )\mathbb{I}_1],\nonumber\\
&k\cdot k\cdot I^{(2)}=\int \frac{d^3q}{(2\pi)^3}[{k^0}^2\mathbb{I}_2-2 k^0(\vec{k}\cdot\vec{q})\mathbb{I}_1+(\vec{k}\cdot\vec{q})^2\mathbb{I}_0],\quad P\cdot P\cdot I^{(2)}=s\int \frac{d^3q}{(2\pi)^3}\mathbb{I}_2,\nonumber\\
&k\cdot I^{(3)}=\int\frac{d^3q}{(2\pi)^3}[k^0\mathbb{I}_3-(\vec{k}\cdot\vec{q})\,\mathbb{I}_2-\vec{q}^{\,2}k^0\mathbb{I}_1+\vec{q}^{\,2}(\vec{k}\cdot\vec{q})\,\mathbb{I}_0],\quad P\cdot I^{(3)}=\sqrt{s}\int\frac{d^3q}{(2\pi)^3}[\mathbb{I}_3-\vec{q}^{\,2}\,\mathbb{I}_1],\label{Intsc2}
\end{align}
with $\sqrt{s}=M_{\phi_R}$ and
\begin{align}
k^0=\frac{s+k^2-p^2}{2\sqrt{s}}.
\end{align}
In Eq.~(\ref{Intsc2}), we have introduced 
\begin{align}
\mathbb{I}_n&\equiv\int\frac{dq^0}{(2\pi)}\frac{{q^0}^n}{\mathcal{D}}=-i\frac{\mathbb{N}_n}{\mathbb{D}}, \label{IIn}
\end{align}
where
\begin{align}
\mathbb{D}&=2E_1 E_2 E_3(\sqrt{s}+E_1+E_2)(k^0+E_2+E_3)(\sqrt{s}-E_1-E_2+i\epsilon)\nonumber\\
&\quad\times(\sqrt{s}-k^0-E_1-E_3+i\epsilon)(-\sqrt{s}+k^0-E_1-E_3+i\epsilon)(k^0-E_2-E_3+i\epsilon),\label{Dbb}
\end{align}
with
\begin{align}
E_1=\sqrt{(\vec{k}+\vec{q}\,)^2+m^2_1},\quad E_2=\sqrt{(\vec{k}+\vec{q}\,)^2+m^2_2}, \quad E_3=\sqrt{\vec{q}^{\,2}+m^2_3}.\label{E123}
\end{align}
For the processes depicted in Fig.~\ref{decay}, $m_1=M_{f_0}$, $m_2=M_\phi$ and $m_3=M_K$. The expressions for $\mathbb{N}_n$ in Eq.~(\ref{IIn}) are,
\begin{align}
\mathbb{N}_0&=(E_1+E_2)\Big[(E_1+E_3)(E_2+E_3)(E_1+E_2+E_3)-E_3 {k^0}^2\Big]\nonumber\\
&\quad-s E_1(E_2+E_3)+2E_1E_3k^0\sqrt{s},\nonumber
\end{align}
\begin{align}
\mathbb{N}_1&=-E_3\Big[k^0(E_1+E_2)\left\{E_1(E_1+E_2+2E_3)+(E_2+E_3)^2-{k^0}^2\right\}\nonumber\\
&\quad+\sqrt{s}\left\{{k^0}^2(2E_1+E_2)-E_2(E_2+E_3)(2E_1+E_2+E_3)\right\}-s k^0 E_1\Big],\nonumber
\end{align}
\begin{align}
\mathbb{N}_2&=E_3\Big[E^3_1\left\{{k^0}^2-E_2(E_2+E_3)\right\}+E^2_1\left\{{k^0}^2(E_2+2E_3)-E_2(E_2+E_3)^2\right\}\nonumber\\
&\quad+E_1\Big\{-E^3_2E_3+\Big(E^2_2-{k^0}^2\Big)\Big((k^0-\sqrt{s})^2-E^2_3\Big)+E_2E_3(\sqrt{s}-2k^0)^2\Big\}\nonumber\\
&\quad+E_2(k^0-\sqrt{s})^2(E_2+E_3-k^0)(E_2+E_3+k^0)\Big],\nonumber
\end{align}
\begin{align}
\mathbb{N}_3&=-E_3\Big[E^3_1 k^0\left\{{k^0}^2-E_2(E_2+2E_3)\right\}+E^2_1\Big\{{k^0}^3(E_2+2E_3)\nonumber\\
&\quad+\sqrt{s}E_2(E_2+E_3-k^0)(E_2+E_3+k^0)-E_2 k^0(E_2+E_3)(E_2+3E_3)\Big\}\nonumber\\
&\quad+E_1\Big\{{k^0}^3(E^2_2+4E_2E_3+E^2_3)+s k^0(E^2_2+2E_2E_3-{k^0}^2)\nonumber\\
&\quad+2\sqrt{s}\Big(E^2_2 E_3(E_2+E_3)-E_2{k^0}^2(E_2+3E_3)+{k^0}^4\Big)\nonumber\\
&\quad-E^2_2 E_3 k^0(2E_2+3E_3)-{k^0}^5\Big\}+E_2(k^0-\sqrt{s})^3(E_2+E_3-k^0)\nonumber\\
&\quad\times(E_2+E_3+k^0)\Big].
\end{align}
To implement in the formalism the unstable character of the $f_0(980)$ resonance, in Eq.~(\ref{Dbb}), the term $-E_1+i\epsilon$ [for the processes studied, $E_1$ is the energy of $f_0(980)$] is replaced by $-E_1+i\frac{\Gamma_{f_0}}{2}$. 

Since we consider an approach in which the states $\phi(2170)$, $K(1460)$, $f_0(980)$ and $K_1(1270)$ are composite hadrons, a form factor is associated with each of the three-vertices involved in the loop in Fig.~\ref{decay}.
In this way, in Eq.~(\ref{Intsc2})
\begin{align}
\int d^3q\to (2\pi) \int\limits_0^\infty d|\vec{q}||\vec{q}|^2\int\limits_{-1}^1d\text{cos}\theta\prod\limits_{i=1}^3F_i(|\vec{q}^{\,*}_i|,\Lambda_i),\label{d3q}
\end{align}
where $\theta$ is the angle between the vectors $\vec{q}$ and $\vec{k}$. For a given decay process (depicted in Fig.~\ref{decay}), the index $i=1,\,2,\,3$ in Eq.~(\ref{d3q}) indicates the three vertices involved in the decay mechanism of $\phi(2170)$, $|\vec{q}^{\,*}_i|$ represents the modulus of the momentum in the center of mass of the vertex $i$ [$\vec{q}$  and $\vec{q}^{\,*}$ in Eq.~(\ref{d3q}) are related through a Lorentz boost] and $\Lambda_i$ are as defined in Refs.~\cite{MartinezTorres:2008gy,Torres:2011jt,Oller:1997ti,Geng:2006yb} ($\Lambda_{\phi_R\to \phi f_0}\sim 2000$ MeV, $\Lambda_{K_R\to K f_0}\sim 1400$ MeV, $\Lambda_{f_0\to K\bar K}\sim 1000$ MeV, $\Lambda_{K^+_1(1270)\to \phi K^+}\sim 750$ MeV). The function $F_i$ in Eq.~(\ref{d3q}) represents the form factor considered for the vertex $i$. In case of regularizing the $d^3q$ integral with a sharp cut-off, a Heaviside $\Theta$-function, i.e.,
\begin{align}
F_i=\Theta(|\vec{q}^{\,*}_i|-\Lambda_i),
\end{align}
is used. A monopole form, i.e.,
\begin{align}
F_i=\frac{\bar{\Lambda}^2_i}{\bar{\Lambda}^2_i+|\vec{q}^{\,*}_i|^2},
\end{align}
or an exponential dependence of the type
\begin{align}
F_i=e^{-\frac{|\vec{q}^{\,*}_i|^2}{2\bar{\Lambda}^2_i}},
\end{align}
are also commonly used as form factors for the vertices. The value of $\bar{\Lambda}_i$, which is similar to the value of $\Lambda_i$, is chosen in such a way that the area under the curve of $F^2_i$ as a function of the modulus of the momentum is same, independently of the form factor used~\cite{Gamermann:2009uq}.

We are now in a position  to calculate $\overline{\sum\limits_\text{pol}}|t|^2$ for the decay processes in Fig.~\ref{decay}, which depends on the coupling constants determined in the previous sections, and evaluate the corresponding decay widths using Eq.~(\ref{wphiR}).

\section{Results}
In this section we present the results obtained for the decay widths of $\phi(2170)$ to a final state involving the kaonic resonances $K^+(1460)$, $K^+_1(1400)$, $K^+_1(1270)$. We will also present the related branching fractions and compare them with the information available from Ref.~\cite{Ablikim:2020pgw}.
\subsection{Decay widths}
In Tables~\ref{T1}-\ref{T3} we show the results obtained for the decay widths of $\phi(2170)\to K^+(1460) K^-$, $K^+_1(1400) K^-$ and $K^+_1(1270) K^-$, respectively. As can be seen, the results determined with different form factors are compatible with each other.  In case of the decay width of $\phi(2170)\to  K^+(1460) K^-$ (see Table~\ref{T1}) we find a value around $0.8-2.0$ MeV. 
\begin{table}[h!]
\caption{Partial decay width (in MeV) of $\phi(2170)\to  K^+(1460)K^-$ by considering different form factors, as explained in Sec.~\ref{For}.}\label{T1}
\begin{tabular}{cc}
\hline
Form factor&Decay width\\
\hline\hline
Heaviside-$\Theta$&$1.5\pm0.5$\\
Monopole&$1.3\pm0.4$\\
Exponential&$1.3\pm0.5$\\
\hline
\end{tabular}
\end{table}

For the decay width of the process $\phi(2170)\to K^+_1(1400) K^-$ (see Table~\ref{T2}), the result found depends on the model considered to determine the coupling of $K^+_1(1400)\to \phi K^+$: within model B, which relates  $K_1(1400)$ and $K_1(1270)$ through a mixing angle, the decay width obtained for $\phi(2170)\to K^+_1(1400) K^-$ is around $1.5-3.1$ MeV. 
\begin{table}[h!]
\caption{Partial decay width (in MeV) of $\phi(2170)\to K^+_1(1400) K^-$ taking into account the different form factors and the models B and C discussed in Sec.~\ref{For} to describe the properties of $K_1(1400)$.}\label{T2}
\begin{tabular}{ccc}
\hline
Form factor&\multicolumn{2}{c}{Decay width}\\
\hline
&Model B&Model C\\
\hline\hline
Heavise-$\Theta$&$\quad2.6\pm0.5\quad$&$15\pm4$\\
Monopole&$\quad1.9\pm0.4\quad$&$11\pm3$\\
Exponential&$\quad2.1\pm0.4\quad$&$12\pm3$\\
\hline
\end{tabular}
\end{table}
However, if we determine the $K^+_1(1400)\to \phi K^+$ coupling considering model C, which uses the data from Ref.~\cite{PDG}, the result obtained for this decay width is $\sim 8-19$ MeV, representing in this way a sizeable contribution to the full width of $\phi(2170)$.  Although it should be reiterated that the experimental data on the radiative decay of $K^+_1(1270)$ and $K^+_1(1400)$ are obtained, through the Primakoff effect, by assuming them as mixture of states belonging to the axial nonets. Thus, the results on the radiative decays in Ref.~\cite{PDG}, and, consequently, the decay width of $\phi(2170)\to  K^+_1(1400) K^-$ found within model C, may need to be taken with caution. We do not discuss the decay of $\phi(2170)\to K_1(1400)  \bar K$ within model A, which treats $K_1(1270)$ as a meson-meson resonance~\cite{Roca:2005nm,Geng:2006yb}, since $K_1(1400)$ was not found to arise from hadron dynamics in these latter works.

For the decay width of $\phi(2170)\to K^+_1(1270)K^- $ (see Table~\ref{T3}), we find that  the result depends on the model used to calculate the coupling of $K^+_1(1270)\to \phi K^+$: within model A, where $K^+_1(1270)$ is generated from vector-pseudoscalar channels and has a double pole structure, the decay width obtained is around $1-2$ MeV when considering the superposition of the two poles. 
\begin{table}[h!]
\caption{Partial decay width (in MeV) of $\phi(2170)\to K^+_1(1270) K^-$ by considering different form factors and the models A, B, C to describe the properties of $K_1(1270)$, as explained in Sec.~\ref{For}.\\}\label{T3}
\begin{tabular}{cccccccc}
\hline
Form factor&\multicolumn{7}{c}{Decay width}\\
\hline
&\multicolumn{3}{c}{Model A}&Model B&\multicolumn{3}{c}{Model C}\\
&Poles $z_1$, $z_2$&Pole $z_1$&Pole $z_2$&&Solution $\mathbb{S}_1$& Solution $\mathbb{S}_2$& Solution $\mathbb{S}_3$\\
\hline\hline
Heaviside-$\Theta$&$\quad1.5\pm0.3\quad$&$0.6\pm0.1\quad$&$0.22\pm0.04\quad$&$0.12\pm0.04\quad$&$1.6\pm0.4\quad$&$17\pm3$&$41\pm9$\\
Monopole&$\quad0.8\pm0.2\quad$&$0.3\pm0.1\quad$&$0.12\pm0.02\quad$&$0.07\pm0.02\quad$&$0.9\pm0.2\quad$&$9\pm2$&$23\pm5$\\
Exponential&$\quad1.0\pm0.2\quad$&$0.4\pm0.1\quad$&$0.15\pm0.03\quad$&$0.09\pm0.02\quad$&$1.1\pm0.3\quad$&$11\pm2$&$28\pm6$\\
\hline
\end{tabular}
\end{table}
Such a  superposition has been implemented in two ways: (1) We use an average mass for $K_1(1270)$ in Eq.~(\ref{tK1}) and the coupling $g_{K^+_1\to\phi K^+}$ is substituted by the sum of the couplings related to the two poles, i.e., $g^{(1)}_{K^+_1(1270)\to\phi K^+}+g^{(2)}_{K^+_1(1270)\to\phi K^+}$. (2) The amplitude $t_{\phi_R\to K^+_1 K^-}$ is written as $t^{(1)}_{\phi_R\to K^+_1 K^-}+t^{(2)}_{\phi_R\to K^+_1 K^-}$, where the superscript indicates the contribution related to each of the two poles. Then the term 2Re$\left\{t^{(1)}_{\phi_R\to K^+_1 K^-}t^{(2)*}_{\phi_R\to K^+_1 K^-}\right\}$ needed to calculate the modulus squared is obtained by using an average mass for $K_1(1270)$. In both cases, an average mass of $K_1(1270)$ is used in the phase space. The results obtained in the two ways are compatible within the uncertainties shown in Table~\ref{T3}.

Continuing with the discussions on the results obtained within the model A, considering the description of Refs.~\cite{Roca:2005nm,Geng:2006yb} for $K_1(1270)$, the contribution to the decay $\phi(2170)\to K^+_1(1270)K^-$ from the pole $z_1$ is larger than the one from the pole $z_2$. This finding is in line with the fact that the former pole couples more to $\pi K^*(892)$~\cite{Roca:2005nm,Geng:2006yb}. It should be mentioned here that of the two poles found in Refs.~\cite{Roca:2005nm,Geng:2006yb} [see Eq.~(\ref{z12})], the mass related to the pole $z_2$ is closer to the value determined from the fit to the experimental data in Ref.~\cite{Ablikim:2020pgw}. However, the process $K^+_1(1270) \to \pi K^*(892)$ is considered in Ref.~\cite{Ablikim:2020pgw}, where the final state couples rather more strongly to the pole $z_1$. Thus, when comparing our results with the experimental information, as we present in the subsequent paragraphs, it might be more meaningful to consider the decay widths obtained from the superposition of the two poles.
In any case, if the two pole nature of $K_1(1270)$ is confirmed, the results in Ref.~\cite{Ablikim:2020pgw} on the related process may require a revision. 

Within the mixing scheme of model B, we find that the results obtained for the decay width of $\phi(2170)\to K^+_1(1270)K^-$ are similar to the ones calculated with model A for the pole $z_2$. Such a result could be in line with the fact that the mass of $K_1(1270)$ in model B is very similar to the mass value associated with the pole $z_2$ in model A. 

Interestingly, if we consider model C, where we used the experimental data available in Ref.~\cite{PDG} to estimate the couplings of $K^+_1(1270)$ and $K^+_1(1400)$ to the $\phi K^+$ channel, we find two different scenarios for the decay width of $\phi(2170)\to  K^+_1(1270)K^-$. In one of them, which corresponds to using solution $\mathbb{S}_1$ of Eq.~(\ref{gK1phen}), the results are compatible with those found in the model A. In the second scenario, which uses solutions $\mathbb{S}_2$ or $\mathbb{S}_3$ of Eq.~(\ref{gK1phen}), a much bigger decay width for $\phi(2170)\to  K^+_1(1270)K^-$ is obtained, which would constitute a sizeable part of the total width of $\phi(2170)$.

\subsection{Branching ratios}
In Ref.~\cite{Ablikim:2020pgw}, the partial decay widths of $\phi(2170)\to  K^+(1460)K^-$, $ K^+_1(1400)K^-$, $K^+_1(1270)K^- $ were not measured. Instead, the products 
$\mathcal{B}r\Gamma^{e^+e^-}_R$, with $\Gamma^{e^+e^-}_R$ being the partial decay width of $\phi(2170)\to e^+ e^-$ and $\mathcal{B}r$  the branching fraction for each of the $\phi(2170)\to R K^-$ processes, with $R=K^+(1460)$, $K^+_1(1400)$, $ K^+_1(1270)$, were extracted. Since the decay width $\Gamma^{e^+e^-}_R$ is not known, we can use the information provided in Ref.~\cite{Ablikim:2020pgw} to calculate the ratios
\begin{align}
B_1\equiv\frac{\Gamma_{\phi_R\to K^+(1460)K^-}}{\Gamma_{\phi_R\to K^+_1(1400)K^-}}=\frac{\mathcal{B}r[\phi_R\to K^+(1460)K^-]}{\mathcal{B}r[\phi_R\to K^+_1(1400)K^-]},\label{Br1}
\end{align}
\begin{align}
B_2\equiv\frac{\Gamma_{\phi_R\to K^+(1460) K^-}}{\Gamma_{\phi_R\to K^+_1(1270)K^-}}=\frac{\mathcal{B}r[\phi_R\to K^+(1460) K^-]}{\mathcal{B}r[\phi_R\to K^+_1(1270)K^-]},\label{Br2}
\end{align}
\begin{align}
B_3\equiv\frac{\Gamma_{\phi_R\to K^+_1(1270) K^-}}{\Gamma_{\phi_R\to K^+_1(1400)K^-}}=\frac{\mathcal{B}r[\phi_R\to K^+_1(1270) K^-]}{\mathcal{B}r[\phi_R\to K^+_1(1400)K^-]},\label{Br3}
\end{align}
and compare with our results. Note that although the above ratios do not depend on the coupling $g_{\phi_R\to \phi f_0}$, the triangular loops and the other vertices involved in the calculation of the decay widths appearing in Eqs.~(\ref{Br1})-(\ref{Br3}) depend on the consideration of $\phi(2170)$ as a $\phi f_0(980)$ state. Thus, the particular values found for the $B_1$, $B_2$ and $B_3$ ratios are related to the nature, not only of $\phi(2170)$, but also to the one of $K(1460)$, $K^+_1(1270)$ and $K^+_1(1400)$.

In Ref.~\cite{Ablikim:2020pgw}, the values (in eV) for the products $\mathcal{B}r\Gamma^{e^+e^-}_R$ are
\begin{align}
\mathcal{B}r[\phi_R\to K^+(1460)K^- ]\Gamma^{e^+e^-}_R&=3.0\pm 3.8,\nonumber\\
\mathcal{B}r[\phi_R\to K^+_1(1400)K^- ]\Gamma^{e^+e^-}_R&=\left\{\begin{array}{c}4.7\pm3.3,~\text{Solution 1}\\98.8\pm7.8,~\text{Solution 2}\end{array}\right.,\nonumber\\
\mathcal{B}r[\phi_R\to  K^+_1(1270)K^-]\Gamma^{e^+e^-}_R&=\left\{\begin{array}{c}7.6\pm3.7,~\text{Solution 1}\\152.6\pm14.2,~\text{Solution 2}\end{array}\right.,\label{Brexp}
\end{align}
having two possible solutions in case of the processes $\phi(2170)\to K^+_1(1400)K^- $, $K^+_1(1270)K^- $ from the fits to the data. Using Eq.~(\ref{Brexp}), we can determine the experimental values for the $B_1$, $B_2$ and $B_3$ ratios, finding
\begin{align}
B^\text{exp}_1&=\left\{\begin{array}{l}0.64\pm0.92,~\text{Solution 1,}\\0.03\pm 0.04,~\text{Solution 2,}\end{array}\right.\nonumber\\
B^\text{exp}_2&=\left\{\begin{array}{l}0.40\pm0.54,~\text{Solution 1,}\\0.02\pm 0.03,~\text{Solution 2,}\end{array}\right.\nonumber\\
B^\text{exp}_3&=\left\{\begin{array}{l}1.62\pm1.38,~\text{Solution 1,}\\1.55\pm 0.19,~\text{Solution 2.}\end{array}\right.\label{Bexp}
\end{align}
Considering now the decay widths listed in Tables~\ref{T1}-\ref{T3}, we can calculate the ratios in Eqs.~(\ref{Br1}),~(\ref{Br2}),~(\ref{Br3}). We present the results in Tables~\ref{TB1}-\ref{TB3}. Since the decay widths obtained in this work do not depend much on the form factors considered, the values presented for the ratios correspond to the average of the results obtained with different form factors.

\begin{table}[h!]
\caption{Results for the branching ratio $B_1$. The label ``Experiment'' refers to the values given in Eq.~(\ref{Bexp}).}\label{TB1}
\begin{tabular}{ccc}
\hline
&&$B_1$\\
\hline\hline
\multirow{2}{*}{Our results}&Model B&$0.62\pm0.20$\\
&Model C&$0.11\pm0.04$\\
\hline
\multirow{2}{*}{Experiment}&Solution 1&$0.64\pm0.92$\\
&Solution 2&$0.03\pm0.04$\\
\hline
\end{tabular}
\end{table}
The ratio $B_1$ [see Eq.~(\ref{Br1})] involves the decay width of $\phi(2170)\to K^+_1(1400) K^-$, thus, it can be calculated within the models B and C. The results obtained in the former case are compatible with the experimental value related to solution 1, while the results in the latter case are closer to the experimental value obtained from solution 2. Although the results obtained in model C can also be compatible with the value found from solution 1 due to the uncertainty present in the experimental data. 
\begin{table}[h!]
\caption{Results for the ratio $B_2$. The label ``Experiment'' refers to the values given in Eq.~(\ref{Bexp}).}\label{TB2}
\begin{tabular}{cccl}
\hline
&&$B_2$\\
\hline\hline
\multirow{7}{*}{Our results}&\multirow{3}{*}{Model A}& $1.3\pm0.4$&(Poles $z_1$, $z_2)$\\
& &$3.6\pm1.2$&(Pole $z_1$)\\
& &$8.8\pm2.8$&(Pole $z_2$)\\
& Model B& $16\pm6$\\
& \multirow{3}{*}{Model C}& $1.2\pm0.4$&(Solution $\mathbb{S}_1$)\\
&  & $0.12\pm0.04$&(Solution $\mathbb{S}_2$)\\
& & $0.05\pm0.02$&(Solution $\mathbb{S}_3$)\\
\hline
\multirow{2}{*}{Experiment}&Solution 1&$0.40\pm0.54$&\\
&Solution 2&$0.02\pm0.03$&\\
\hline
\end{tabular}
\end{table}

As can be seen from Table~\ref{TB2}, the value of $B_2$ depends on the description considered for $K^+_1(1270)$. Within model A [in this case, $K_1(1270)$ has a double pole structure], we find that the interference between the two poles leads to a value which is closer to the upper limit for this ratio obtained with solution 1 of the BESIII Collaboration. We also find that the contribution from the individual poles of $K^+_1(1270)$ produces a larger value for $B_2$, which is not compatible with the experimental value.  In the model  B, the values obtained for $B_2$ are not compatible with those determined from the experimental data. In case of using model C, solutions $\mathbb{S}_2$ and $\mathbb{S}_3$ give rise to a value for $B_2$ which is compatible with solution 2 of Ref.~\cite{Ablikim:2020pgw}. Solution $\mathbb{S}_1$, instead, produces a value for $B_2$ which is compatible with solution 1 of Ref.~\cite{Ablikim:2020pgw}. 

\begin{table}[h!]
\caption{Results for the ratio $B_3$. The label ``Experiment'' refers to the values given in Eq.~(\ref{Bexp}).}\label{TB3}
\begin{tabular}{cccl}
\hline
&&$B_3$\\
\hline\hline
\multirow{4}{*}{Our results}& Model B& $0.04\pm0.01$\\
& \multirow{3}{*}{Model C}& $0.09\pm0.02$&(Solution $\mathbb{S}_1$)\\
&  & $0.96\pm0.16$&(Solution $\mathbb{S}_2$)\\
& & $2.40\pm0.40$&(Solution $\mathbb{S}_3$)\\
\hline
\multirow{2}{*}{Experiment}&Solution 1&$1.62\pm1.38$&\\
&Solution 2&$1.55\pm0.19$&\\
\hline
\end{tabular}\end{table}

The results for the ratio $B_3$ can be found in Table~\ref{TB3}. Since this ratio involves the decay width of $\phi(2170)\to K^+_1(1400)K^- $, we evaluate it within models B and C. Although, due to the similarity between the decay width for $\phi(2170)\to K^+_1(1270) K^-$ within model A (considering the superposition of two poles for $K_1(1270)$) and solution $\mathbb{S}_1$ of model C, it can be inferred that the ratio $B_3$ (under solution $\mathbb{S}_1$ in Table~\ref{TB3}) represent the result for both cases. It can be said, then, that for solution $\mathbb{S}_1$, as well as for model A, the results can be considered to be closer to the lower limit of solution 1 presented in Table~\ref{TB3}. Solutions $\mathbb{S}_2$ and $\mathbb{S}_3$ of model C are compatible with the data. 

To summarize the findings of the present work, we can state:
\begin{itemize}
\item{The $\phi f_0$ description of $\phi(2170)$ can straightforwardly explain its suppressed decay to $\bar K^*(892) K^*(892)$, which is one of the findings of the BESIII Collaboration.}

\item{A branching ratio $B_1$ [defined in Eq.~(\ref{Br1})] for the $\phi(2170)$ decay to final states involving $K(1460)$ and $K_1(1400)$ is calculated treating the former as a $Kf_0$ state and the latter within two different models. One of the models (model B) relates $K_1(1270)$ and $K_1(1400)$ through a mixing angle~\cite{Palomar:2003rb}, while the other one (model C) is based on a phenomenological determination of the $K_1(1400)\phi K$ coupling using the information available on its hadronic and radiative decays. The results obtained within both models are compatible with the ratio evaluated using experimental data.}

\item{A ratio $B_2$ [defined in Eq.~(\ref{Br2})] for the $\phi(2170)$ decay to final states involving $K(1460)$ and $K_1(1270)$ is obtained using yet another model (model A) for the latter one, besides the two mentioned in the previous point. Within model A, $K_1(1270)$ is interpreted as a state, related to two poles in the complex energy plane, arising from pseudoscalar-vector meson dynamics. The ratio $B_2$ obtained within model A is in agreement with the data, when the superposition of the two poles is considered. The former result is found to be similar to that obtained within a phenomenological description for $K_1(1270)$ and $K_1(1400)$ (solution $\mathbb S_1$ of model C), which may indicate that the information on the superposition of the two poles is present in the experimental data used to obtain the phenomenological solution. The ratio $B_2$ does not get reproduced within model $B$.}

\item{A third ratio, $B_3$ [defined in Eq.~(\ref{Br3})], for the $\phi(2170)$ decay to final states involving $K_1(1400)$ and $K_1(1270)$ is calculated using models B and C. This ratio is in agreement with the values obtained from the experimental data when using model C (solutions $\mathbb S_2$ and $\mathbb S_3$). The ratio obtained using the solution $\mathbb S_1$ of model C too (and, hence, within model A, in which case the decay width to $K^+_1(1270)K^-$ is similar) is also close to the lower limit of the value determined from the data (based on solution 1 in Ref.~\cite{Ablikim:2020pgw}).} 

\item{It can be said that the $\phi f_0$ description of $\phi(2170)$ can well describe the experimental findings of Ref.~\cite{Ablikim:2020pgw}. The moleculelike nature, related to two poles arising from meson-meson dynamics, and a phenomenological description (solution $\mathbb S_1$ of model C) of $K_1(1270)$ seem to be in agreement. A model relating $K_1(1270)$ and $K_1(1400)$ through a mixing angle as in Ref.~\cite{Palomar:2003rb}, does not describe two of the three-ratios, indicating that a different mixing scheme may be required for such a relation.}  

\end{itemize}

\section{Conclusions}
In this work we have obtained the decay widths of $\phi(2170)$ to $K^+(1460)K^- $, $K^+_1(1400) K^-$ and $K^+_1(1270)K^-$ within an approach in which $\phi(2170)$ is interpreted as a $\phi f_0(980)$ molecular state and $K^+(1460)$ as a state originated from the $Kf_0(980)$ interaction. In case of $K^+_1(1270)$ and $K^+_1(1400)$ we have used different models to describe their properties. Considering the decay widths determined, we calculate the ratios $B_1=\frac{\Gamma_{\phi(2170)\to K^+(1460)K^-}}{\Gamma_{\phi(2170)\to K^+_1(1400)K^-}}$, $B_2=\frac{\Gamma_{\phi(2170)\to K^+(1460) K^-}}{\Gamma_{\phi(2170)\to K^+_1(1270)K^-}}$ and $B_3=\frac{\Gamma_{\phi(2170)\to K^+_1(1270) K^-}}{\Gamma_{\phi(2170)\to K^+_1(1400)K^-}}$ and compare with the corresponding values found from the experimental data on $\mathcal{B}r\Gamma^{e^+e^-}_R$ of Ref.~\cite{Ablikim:2020pgw}. We obtain results for these ratios which are compatible with the latter ones. Further experimental data with higher statistics can be very helpful in drawing more robust conclusions on the properties of $K_1(1270)$ and $K_1(1400)$. The partial decay widths provided in the present work can be useful for future experimental investigations.

\section{Acknowledgements}
 This work is supported by the Funda\c c\~ao de Amparo \`a Pesquisa do Estado de S\~ao Paulo (FAPESP), processos n${}^\circ$ 2019/17149-3, 2019/16924-3 and 2020/00676-8, by the Conselho Nacional de Desenvolvimento Cient\'ifico e Tecnol\'ogico (CNPq), grant  n${}^\circ$ 305526/2019-7 and 303945/2019-2 and by the Deutsche Forschungsgemeinschaft (DFG, German Research Foundation), in part through the Collaborative Research Center [The Low-Energy Frontier of the Standard Model, Project No. 204404729?SFB 1044], and in part through the Cluster of Excellence [Precision Physics, Fundamental Interactions, and Structure of Matter] (PRISMA${}^+$ EXC 2118/1) within the German Excellence Strategy (Project ID 39083149).
 
\appendix
\section*{Appendices}
\section{Model for the process $e^+e^-\to \phi f_0(980)$}\label{AphiR}
\begin{figure}[h!]
\centering
\includegraphics[width=0.45\textwidth]{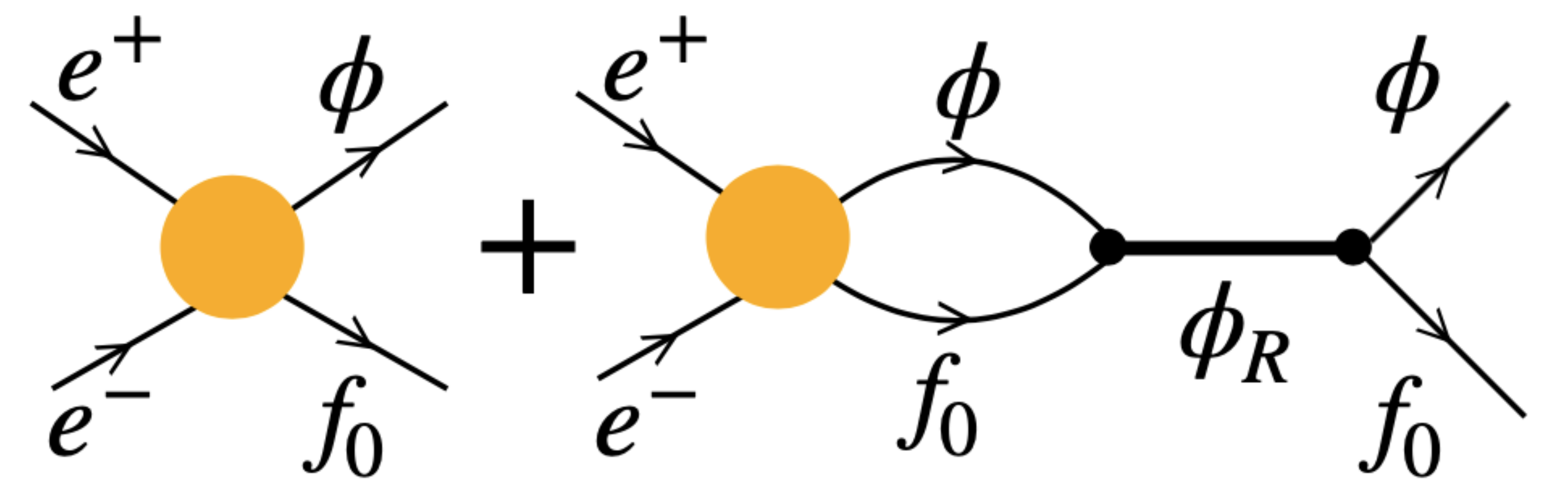}
\caption{Diagrammatic description of the process $e^+e^-\to\phi f_0$ with the formation of $\phi(2170)$ from the interaction of $\phi f_0$ as in Ref.~\cite{MartinezTorres:2008gy}.(Left) Tree level contribution to the process, where a $\phi$ and a $f_0$ is produced in the final state as plane waves. (Right) The production of $\phi f_0$, followed by the final state interactions leading to the formation (and decay) of $\phi(2170)$.}\label{phiR}
\end{figure}

Within our description of $\phi(2170)$ as a $\phi f_0$ molecular state, the formation of $\phi(2170)$ in the process $e^+ e^-\to \phi f_0(980)$ proceeds as shown in Fig.~\ref{phiR} [depicting the tree-level contribution and the one  with the final state interactions forming $\phi(2170)$]. 
At the tree level, $e^+$ and $e^-$ interact and produce  a $\phi$ and  a $f_0(980)$ as  plane waves. In Ref.~\cite{MartinezTorres:2008gy}, such a background contribution was described by using the results obtained in Ref.~\cite{Napsuciale:2007wp}. 
Then, the $\phi$ and $f_0(980)$ propagate and interact in the final state, forming $\phi(2170)$, which, subsequently, decays into $\phi$ and $f_0(980)$. In this way, the amplitude for the the process $e^+e^-\to \phi f_0(980)$ can be obtained by multiplying the non-resonant contribution or the background by the factor $|1+G_{\phi f_0}T_{\phi f_0\to \phi f_0}|^2$, where $G_{\phi f_0}$ is the loop function for the virtual $\phi f_0(980)$ state (a cut-off of the order $M_\phi+M_{f_0}$ is used to regularize it).
\section{Evaluation of the decay width for the process $K^+_1\to \phi K^+$}\label{widthK1}
In the tensor formalism of Ref.~\cite{Palomar:2003rb}, the decay width of $K^+_1\to\phi K^+$, $\Gamma^T_{K^+_1\to \phi K^+}$, can be determined as 
\begin{align}
\Gamma^T_{K^+_1\to \phi K^+}&=\frac{|g^T_{K^+_1\to \phi K^+}|^2}{2\pi}\frac{1}{\mathcal{N}}\int\limits_{M_{K_1}-a\Gamma_{K_1}}^{M_{K_1}+a\Gamma_{K_1}} d\tilde{M}_{K_1}(2\tilde{M}_{K_1})\frac{|\vec{p}|}{\tilde{M}^2_{K_1}}\Big[1+\frac{2}{3}\frac{|\vec{p}|}{M^2_\phi}\Big]\nonumber\\
&\quad\quad\times\text{Im}\Bigg[\frac{1}{\tilde{M}^2_{K_1}-M^2_{K_1}+i M_{K_1}\Gamma_{K_1}}\Bigg]\theta(\tilde{M}_{K_1}-M_\phi-M_K)\theta(\tilde{M}_{K_1}-M_\pi-M_{K^*(892)}),\label{wT}
\end{align}
where we incorporate the effect of the finite width of $K_1$ by convoluting on its mass. Typically, in the integral limits, a value $a\simeq 2-3$ is used to cover the energy region associated with the resonance. The Heaviside $\theta$-functions in Eq.~(\ref{wT}) guarantee energy conservation as well as that $K_1$ has a mass big enough for decaying to its lowest decay channel when convoluting. In Eq.~(\ref{wT}), $|\vec{p}|$ is the modulus of the center of mass momentum, $\mathcal{N}$ is a normalization factor given by
\begin{align}
\mathcal{N}=\int\limits_{M_{K_1}-a\Gamma_{K_1}}^{M_{K_1}+a\Gamma_{K_1}} d\tilde{M}_{K_1} (2\tilde{M}_{K_1})\text{Im}\Bigg[\frac{1}{\tilde{M}^2_{K_1}-M^2_{K_1}+i M_{K_1}\Gamma_{K_1}}\Bigg],
\end{align}
and, from Ref.~\cite{Palomar:2003rb},
\begin{align}
g^T_{K^+_1\to \phi K^+}=\left\{\begin{array}{c}\text{cos}\alpha \tilde{D}+\text{sen}\alpha \tilde{F},~\text{for}~K_1(1270),\\\text{sen}\alpha\tilde{D}-\text{cos}\alpha\tilde{F},~\text{for}~K_1(1400).\end{array}\right.
\end{align}
Using the values of  $\tilde{D}$ and $\tilde{F}$, as a function of the mixing angle,  $\alpha$,  as given in (Table 7 of) Ref.~\cite{Palomar:2003rb}, we get 
\begin{align}
\Gamma^T_{K^+_1(1270)\to \phi K^+}=\left\{\begin{array}{l} 0.030~\text{MeV},~\alpha=29^\circ,\\0.023~\text{MeV},~\alpha=47^\circ,\\0.043~\text{MeV},~\alpha=62^\circ,\end{array}\right.\quad
\Gamma^T_{K^+_1(1400)\to \phi K^+}=\left\{\begin{array}{l} 6.7~\text{MeV},~\alpha=29^\circ,\\6.7~\text{MeV},~\alpha=47^\circ,\\6.6~\text{MeV},~\alpha=62^\circ.\end{array}\right. \label{Tgamval}
\end{align}

Using now Eq.~(\ref{ts}), the decay width of $K^+_1\to \phi K^+$ can be determined within the approach in which vector and axial mesons are described as vector fields instead of second rank tensor fields. In this case, the decay width of $K^+_1\to \phi K^+$ is obtained as
\begin{align}
\Gamma_{K^+_1\to \phi K^+}&=\frac{|g_{K^+_1\to \phi K^+}|^2}{24\pi}\frac{1}{\mathcal{N}}
\int\limits_{M_{K_1}-a\Gamma_{K_1}}^{M_{K_1}+a\Gamma_{K_1}} d\tilde{M}_{K_1}(2\tilde{M}_{K_1})\frac{|\vec{p}|}{\tilde{M}^2_{K_1}}\left[3+\frac{|\vec{p}|^2}{M^2_\phi}\right]\nonumber\\
&\quad\quad\quad \times\text{Im}\Bigg[\frac{1}{\tilde{M}^2_{K_1}-M^2_{K_1}+i M_{K_1}\Gamma_{K_1}}\Bigg]\theta(\tilde{M}_{K_1}-M_\phi-M_K)\theta(\tilde{M}_{K_1}-M_\pi-M_{K^*(892)}).\label{GT}
\end{align}
The value $|g_{K^+_1\to \phi K^+}|$ is determined by equating Eqs.~(\ref{GT}) and (\ref{Tgamval}).

\section{Determination of the $K_1\to \phi K$ coupling within a phenomenological approach}\label{radiative}
\begin{figure}[h!]
\centering
\includegraphics[width=0.3\textwidth]{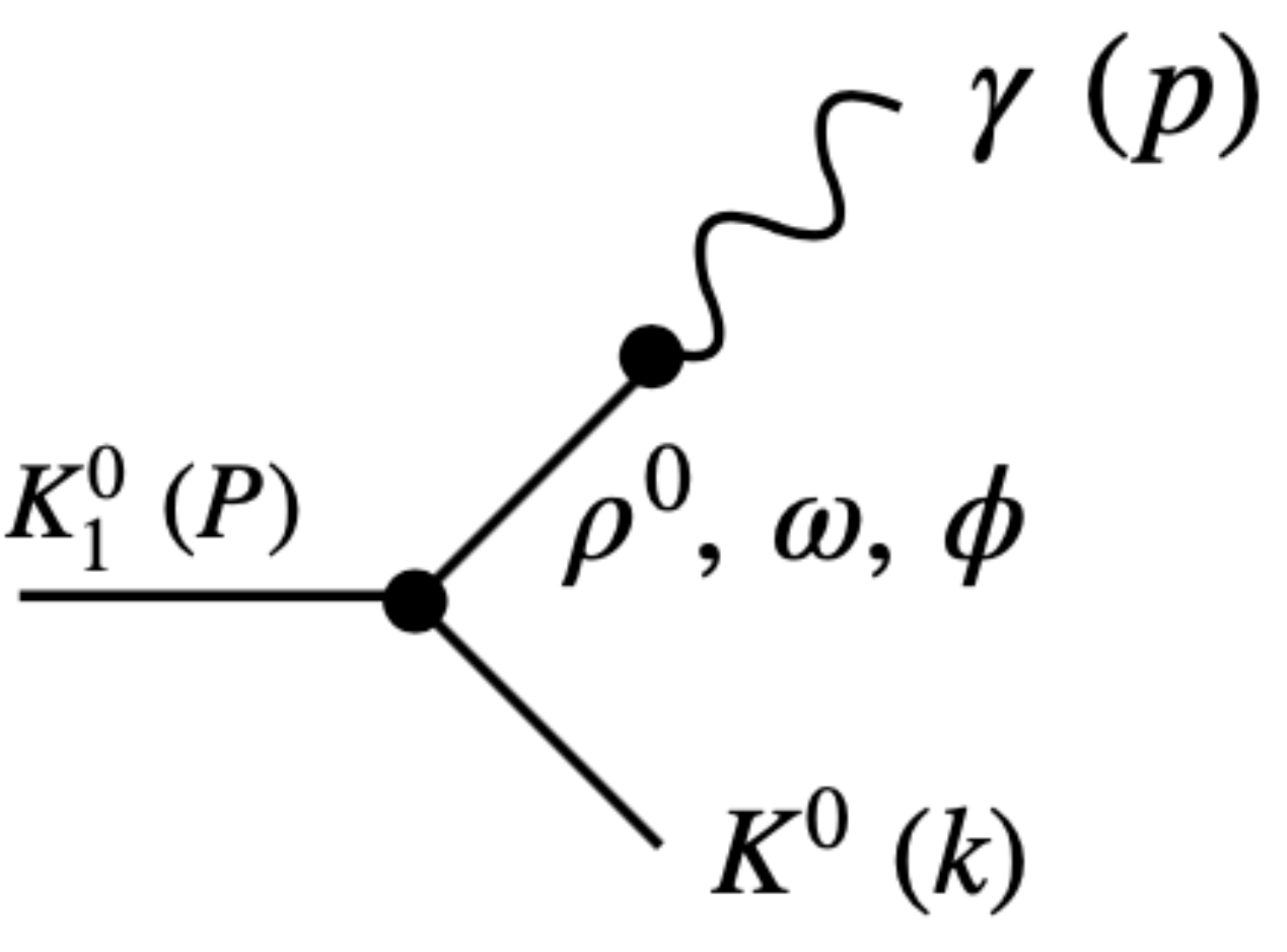}
\caption{Diagrammatic representation of the process $K^0_1\to \gamma K^0$, where $K^0_1$ represents $K^0_1(1270)$ or $K^0_1(1400)$.}\label{K1dec}
\end{figure}

Let us examine how to get the $K_1\phi K$ coupling using the data on radiative and hadronic decays. We start by considering that the radiative decay of $K_1$ proceeds through the vector meson dominance mechanism~\cite{Sakurai:1960ju,Bando:1984ej,Bando:1987br}. In this way, the decay of $K^0_1\to\gamma K^0$ at the tree level can be described as depicted in Fig.~\ref{K1dec}.
Since the decay widths for $K^0_1\to \rho^0 K^0,~\omega K^0$ are known~\cite{PDG}, we can determine $|g_{K^0_1\to \rho^0 K^0}|$ and $|g_{K^0_1\to \omega K^0}|$ and use the information to calculate $|g_{K^0_1\to \phi K^0}|$ such as  to reproduce the known radiative decay width of $K^0_1$. If we use the expression in Eq.~(\ref{ts}) to describe the vertex $K^0_1\to V K^0$, where $V=\rho^0,\,\omega,\,\phi$, the amplitude obtained for the process represented in Fig.~\ref{K1dec} is given by
\begin{align}
t_{K^0_1\to K^0\gamma}=\frac{e M^2_V}{3g}\Bigg[\frac{1}{\sqrt{2}}\left(\frac{\sqrt{3}g_{K^0_1\to \rho^0 K^0} }{M^2_\rho}+\frac{g_{K^0_1\to \omega K^0}}{M^2_\omega}\right)-\frac{g_{K^0_1\to\phi K^0}}{M^2_\phi}\Bigg]\epsilon_{K^0_1}(P)\cdot \epsilon_\gamma(p),\label{trad}
\end{align}
where the Lagrangian~\cite{Bando:1987br}
\begin{align}
\mathcal{L}_{V\to\gamma}=-M^2_V\frac{e}{g}A^\mu\Big(\frac{1}{3\sqrt{2}}\omega_\mu+\frac{1}{\sqrt{2}}\rho^0_\mu-\frac{1}{3}\phi^\mu\Big),
\end{align}
with $A^\mu$ denoting the photon field, $e^2=4\pi\alpha$ ($\alpha$ is the structure constant) and $g=\frac{M_V}{2 f_\pi}$ ($M_V\simeq M_\rho$, $f_\pi\simeq 93$ MeV), has been used for the $V\to \gamma$ transition. As can be seen by replacing $\epsilon_\gamma\to p$, the amplitude in Eq.~(\ref{trad}) is not gauge invariant. An alternative way of determining $g_{K^0_1\to \phi K^0}$ would be to attribute a tensor field to the vector/axial mesons~\cite{Ecker:1988te,Xiong:1992ui}. The amplitude for $K^0_1\to \gamma K^0$ in such a formalism is explicitly gauge invariant. In fact, in the tensor formalism of Refs.~\cite{Ecker:1988te,Xiong:1992ui,Palomar:2003rb},
\begin{align}
t^T_{K^0_1\to \gamma K^0}&=-\frac{2 e F_V}{M_{K^0_1}}\Big[\frac{g^T_{K^0_1\to \rho^0 K^0}}{M^2_{\rho^0}}+\frac{g^T_{K^0_1\to \omega K^0}}{3 M^2_\omega}-\frac{\sqrt{2}g^T_{K^0_1\to\phi K^0}}{3M^2_\phi}\Big]\nonumber\\
&\quad\times\Big[(P\cdot p)(\epsilon_{K^0_1}(P)\cdot \epsilon_\gamma(p))-(P\cdot\epsilon_\gamma(p))(p\cdot\epsilon_{K^0_1}(P))\Big],
\end{align}
with $F_V\simeq 154$ MeV, and, the decay width of $K^0_1\to \gamma K^0$ is given by 
\begin{align}
\Gamma^T_{K^0_1\to \gamma K^0}=\frac{|\vec{p}|^3}{3\pi M^2_{K^0_1}}e^2 |F_V|^2\Bigg|\frac{g^T_{K^0_1\to \rho^0 K^0}}{M^2_{\rho^0}}+\frac{g^T_{K^0_1\to \omega K^0}}{3 M^2_\omega}-\frac{\sqrt{2}g^T_{K^0_1\to\phi K^0}}{3M^2_\phi}\Bigg|^2.\label{GTg}
\end{align}

We now determine the values of $|g^T_{K^0_1\to \rho^0 K^0}|$ and $|g^T_{K^0_1\to \omega K^0}|$ within the tensor formalism such as to reproduce the experimental data on the decay widths of $K_1\to \rho K$ and $K_1\to \omega K$. Let us discuss first the case of $K_1(1270)$. According to Ref.~\cite{PDG}, 
\begin{align}
\Gamma^\text{exp}_{K_1(1270)}&=90\pm 20~\text{MeV},\nonumber\\
\Gamma^\text{exp}_{K_1(1270)\to K\rho}&=(0.42\pm 0.06)\Gamma_{K^\text{exp}_1(1270)},\nonumber\\
\Gamma^\text{exp}_{K_1(1270)\to K\omega}&=(0.11\pm 0.02)\Gamma_{K^\text{exp}_1(1270)},\nonumber\\
\Gamma^\text{exp}_{K^0_1(1270)\to K^0\gamma}&=(73.2\pm 6.1\pm28.3)~\text{KeV}.\label{Gexp}
\end{align}
By using Eq.~(\ref{wT}), substituting $\phi\to \rho,\,\omega$, and by generating random numbers for the known widths for $K_1(1270)\to K\rho, K\omega$ (in the interval allowed by the related error) we can estimate $|g^T_{K^0_1\to \rho^0 K^0}|$ and $|g^T_{K^0_1\to \omega K^0}|$, and find
\begin{align}
|g^T_{K^0_1(1270)\to \rho^0 K^0}|=1104\pm77~\text{MeV},\quad |g^T_{K^0_1(1270)\to \omega K^0}|=1514\pm102~\text{MeV}.\label{gTrho}
\end{align}
In this case, when obtaining $|g^T_{K^0_1\to \rho^0 K^0}|$, we use isospin relations and the width of the $\rho$-meson is taken into account by considering another integral around the nominal mass of $\rho$ in Eq.~(\ref{wT}). Now, by using Eqs.~(\ref{GTg}), (\ref{Gexp}) and (\ref{gTrho}) we can extract the value of $|g^T_{K^0_1(1270)\to\phi K^0}|$ using $\Gamma^\text{exp}_{K^0_1(1270)\to\gamma K^0}$. Here, we must emphasize that only the modulus of $g^T_{K^0_1\to \rho^0 K^0}$ and $g^T_{K^0_1\to \omega K^0}$ can be determined from the experimental data, when, in general, the couplings in Eq.~(\ref{GTg}) can be complex numbers.  We assume them to be real numbers,  which can be either positive or negative. We then generate random numbers inside the interval allowed by the error related to $g^T_{K^0_1\to \rho^0 K^0}$, $g^T_{K^0_1\to \omega K^0}$ [as in Eq.~(\ref{gTrho})] and consider the different sign combinations for the couplings. We then determine the average value and the standard deviation for  $g^T_{K^0_1\to \phi K^0}$.  Independently of the sign chosen for the couplings, we find three different solutions for $|g^T_{K^0_1(1270)\to \phi K^0}|$
\begin{align}
|g^T_{K^0_1(1270)\to \phi K^0}|=\left\{\begin{array}{c}1887\pm 590~\text{MeV},~\text{Solution $\mathbb{S}_1$,}\\~6300\pm 529~\text{MeV},~\text{Solution $\mathbb{S}_2$,}\\~9996\pm 583~\text{MeV},~\text{Solution $\mathbb{S}_3$,}\end{array}\right.
\end{align}
By using now Eq.~(\ref{wT}), we get
\begin{align}
\Gamma^T_{K^0_1(1270)\to \phi K^0}=\left\{\begin{array}{c}0.22\pm0.08~\text{MeV},~\text{Solution $\mathbb{S}_1$},\\2.21\pm 0.46~\text{MeV},~\text{Solution $\mathbb{S}_2$,}\\5.52\pm 1.07~\text{MeV},~\text{Solution $\mathbb{S}_3$}.\end{array}\right.\label{GK1T}
\end{align}
Since the decay width of $K^0_1(1270)\to \phi K^0$ is not known, we consider the three solutions for $|g^T_{K^0_1(1270)\to \phi K^0}|$ as valid and investigate the implications in the calculation of the decay width of $\phi(2170)$. Using the values  in Eq.~(\ref{GK1T}) as input,  we can calculate $|g_{K^0_1\to \phi K^0}|$, which coincides with $|g_{K^+_1\to \phi K^+}|$, related to the amplitudes written by attributing a vector field to the axial/vector mesons. We find
\begin{align}
|g_{K^+_1(1270)\to \phi K^+}|=\left\{\begin{array}{c}3967\pm 419~\text{MeV},~\text{Solution $\mathbb{S}_1$},\\~12577\pm763~\text{MeV},~\text{Solution $\mathbb{S}_2$},\\~19841\pm1177~\text{MeV},~\text{Solution $\mathbb{S}_3$}.\end{array}\right.
\end{align}

We can now repeat the same procedure for $K_1(1400)$ and estimate $|g_{K^0_1(1400)\to \phi K^0}|$. In this case, according to Ref.~\cite{PDG},
\begin{align}
\Gamma^\text{exp}_{K_1(1400)\to \rho K}=1-3~\text{MeV},
\end{align}
while for the decay width of $K_1(1400)\to\omega K$ different experiments have found very different values,
\begin{align}
\Gamma^\text{exp}_{K_1(1400)\to \omega K}=\left\{\begin{array}{l} 11-35~\text{MeV},\\\left(0.01\pm0.01\right)
\Gamma_{K^\text{exp}_1(1400)}.\end{array}\right.
\end{align}

Further, using $\Gamma^\text{exp}_{K^0_1(1400)\to \gamma K^0}=280.8\pm23.2\pm40.4$ KeV~\cite{PDG} and following the same procedure as explained for the determination of the $K_1(1270) \phi K$ coupling, we obtain
\begin{align}
|g_{K^+_1(1400)\to \phi K^+}|=8480\pm 1333~\text{MeV}.\label{gK114A}
\end{align}

\bibliographystyle{unsrt}
\bibliography{refs}

\end{document}